\documentclass[fleqn,usenatbib]{mnras}

\usepackage[T1]{fontenc}
\usepackage{ae,aecompl}

\usepackage{graphicx}	
\usepackage{amsmath}	
\usepackage{amssymb}	

\usepackage{aas_macros}
\usepackage[normalem]{ulem}
\usepackage[usenames]{color}
\usepackage{bm}
\usepackage{cancel}
\usepackage{newtxtext,newtxmath}

\definecolor{DarkGreen}{rgb}{0.64,0.80,0.35}

\title{}
\title[Blazar heating and the Lyman-$\alpha$ forest]{Constraining blazar heating with the $2\lesssim z\lesssim 3$ Lyman-$\alpha$ forest}

\author[A. Lamberts et. al.]{
Astrid Lamberts$^{1,2}$\thanks{E-mail: astrid.lamberts@oca.eu},
Ewald Puchwein$^{3}$,
Christoph Pfrommer$^{3}$,
Philip Chang$^{4}$,
\newauthor Mohamad Shalaby$^{3}$,
Avery Broderick$^{5,6}$,
Paul Tiede$^{5,6}$,
Gwen Rudie$^{7}$
\\
$^{1}$Universit\'e C\^ote d'Azur, Observatoire de la C\^ote d'Azur, CNRS, Laboratoire Lagrange,  Bd de l'Observatoire,\\
CS 34229, 06304 Nice cedex 4, France.\\
$^{2}$Universit\'e C\^ote d'Azur, Observatoire de la C\^ote d'Azur, CNRS, Laboratoire Art\'emis, Bd de l'Observatoire,\\
CS 34229, 06304 Nice cedex 4, France.\\
$^{3}$Leibniz Institute for Astrophysics, Potsdam (AIP), An der Sternwarte 16, D-14482 Potsdam, Germany\\
$^{4}$Department of Physics, University of Wisconsin-Milwaukee, 3135 N. Maryland Ave., Milwaukee, WI 53211, USA\\
$^{5}$Department of Physics and Astronomy, University of Waterloo, 200 University Avenue West, Waterloo, ON, N2L 3G1, Canada\\
$^{6}$Perimeter Institute for Theoretical Physics, 31 Caroline Street North, Waterloo, ON, N2L 2Y5, Canada\\
$^7$The Observatories of the Carnegie Institution for Science, 813 Santa Barbara Street, Pasadena, CA 91101, USA
}

\date{Accepted XXX. Received YYY; in original form ZZZ}

\pubyear{2019}

\begin{document}
\label{firstpage}
\pagerange{\pageref{firstpage}--\pageref{lastpage}}
\maketitle

\begin{abstract}
The intergalactic medium (IGM) acts like a calorimeter recording energy injection by cosmic structure formation, shocks and photoheating from stars and active galactic nuclei. It was recently proposed that spatially inhomogeneous TeV-blazars could significantly heat up the underdense IGM, resulting in patches of both cold and warm IGM around $z\simeq2-3$. The goal of this study is to compare predictions of different blazar heating models with recent observations of the IGM. We perform a set of cosmological simulations and carefully compute mock observables of the Lyman-$\alpha\ ($Ly$\alpha$) forest. We perform a detailed assessment of different systematic uncertainties which typically impact this type of observables and find that they are smaller than the differences between our models. We find that our inhomogeneous blazar heating model is in good agreement with the Ly$\alpha$ line properties and the rescaled flux probability distribution function at high redshift ($2.5<z<3$) but that our blazar heating models are challenged by lower redshift data ($2<z<2.5$). Our results could be explained by HeII reionisation although  state-of-the-art models fall short on providing enough heating to the low-density IGM, thus motivating further radiative transfer studies of inhomogeneous HeII reionisation. 
If blazars are indeed hosted by group-mass halos of $2\times10^{13}\,\rmn{M}_\odot$, a later onset of blazar heating in comparison to previous models would be favoured, which could bring our findings here in agreement with the evidence of blazar heating from local gamma-ray observations.
\end{abstract}

\begin{keywords}
methods: numerical, data analysis - intergalactic medium - quasars: absorption lines
\end{keywords}



\section{Introduction}

Even at the present day, the majority of the baryons reside in the intergalactic medium (IGM) rather than  within galaxies \citep{2012ApJ...759...23S}.  As the main reservoir for baryons, the physical state of the IGM sets the initial conditions for  galaxy formation. Due to its low density, the evolution of the IGM is mostly linear, closely follows the underlying dark matter and is directly influenced by fundamental cosmological parameters \citep{2004MNRAS.354..684V,2013A&A...559A..85P,2013JCAP...04..026S,2015JCAP...11..011P}.  

Because of its linear nature, the IGM is an excellent calorimeter of energy injected by star formation and active galactic nuclei. More specifically, reionisation of H and HeI around  $z\simeq 10-5.3$ \citep{2006ARA&A..44..415F, 2019MNRAS.485L..24K} and of HeII around $4\lesssim z\lesssim 2.7$ provide most of the heat input in the IGM \citep{2009ApJ...694..842M,2011ApJ...733L..24W,2016ApJ...825..144W}. Subsequently, the thermal evolution of the IGM is set by photoheating following recombination and adiabatic cooling due to the Hubble expansion. The lowest density gas expands fastest and also receives the least photoheating as there are fewer recombinations that allow subsequent photoionisations. Together this yields, well after reionisation, a tight temperature-density relation $T=T_0 \Delta ^{\gamma-1}$ for low density gas \citep{1997MNRAS.292...27H,2015MNRAS.450.4081P,2016MNRAS.456...47M} where $T_0$ is the temperature at the mean density, $\Delta=\rho/\bar{\rho}-1$ is the baryon overdensity, and $\gamma$ asymptotically approaches $\simeq 1.6$.  In this paper, we compare recent observational data with different models deviating from a tight power-law $T-\rho$ relation with $\gamma \simeq 1.6$.

During  and shortly after HeII reionisation, the thermal state of the IGM is more complex.  Recent models of HeII reionisation, based on cosmological simulations including complete radiative transfer, indicate a patchy process yielding a broadening of the temperature-density distribution around and below mean density for $z\simeq 3$ \citep{2009ApJ...694..842M,2012MNRAS.423....7M,2013MNRAS.435.3169C,2017ApJ...841...87L}.

Nevertheless, the most dramatic impact on the low-density IGM could come from TeV-blazar heating \citep{2012ApJ...752...23C,2012ApJ...752...24P,Puchwein_12_Lya,2015ApJ...811...19L}. This model is based on the electron/positron beams that result from pair-production from TeV gamma-rays from blazars on the extragalactic background light. 

The pair beams can be subject to plasma instabilities,
 which efficiently convert their energy into plasma modes within the IGM.  If this energy is efficiently thermalised via, e.g., the nonlinear plasma mode interactions, the kinetic energy of the pair beams will ultimately be redistributed to the surrounding IGM \citep{2012ApJ...752...22B,2013ApJ...777...49S,2012ApJ...758..102S,2014ApJ...797..110C,2016ApJ...833..118C,2017ApJ...841...52S,2017ApJ...848...81S,2017A&A...607A.112R,2018ApJ...857...43V},  but see \citet{2013ApJ...770...54M,2014ApJ...787...49S}.  The resulting heating  would only be limited by the number of TeV-photons, which makes it a competitive heating source in underdense regions of the IGM \citep{2012ApJ...752...23C}, where photoheating is  insignificant as the recombination time is larger than the Hubble time.  Assuming a uniform heating rate, blazar heating results in an inverted temperature-density distribution below the cosmic mean ($\gamma \leq 1$), with low density gas reaching  $10^5$K at $z=3$ and  $\simeq 3 \times 10^5$K in the present-day universe \citep[see Fig. 12 in][P12 hereafter]{Puchwein_12_Lya}.

Although uniform heating is a reasonable first order approximation, in \citet[hereafter Paper~I]{2015ApJ...811...19L} we showed that  the heating is affected by the clustering of blazars. As a result, regions close to large overdense regions such as clusters or groups are receiving more heat than remote regions mostly surrounded by voids.\footnote{To ease comparison, we only varied the spatial heating rate in Paper~I and kept the redshift evolution identical. In principle, a more highly biased population should also evolve later in a hierarchically growing universe, i.e., the blazar heating rate should have a different redshift evolution for the differently biased models. We postpone a study of this effect to future work.}  In our favoured model, where TeV blazars have the same bias as quasars, there are almost two orders of magnitude in temperature between the hottest and coldest gas between $z\simeq 2-3$. However, the bulk of the gas follows a temperature consistent with the uniform model. By the present day, all regions would have been heated up and the uniform model provides a good description of the impact of blazar heating. The goal of this paper is to compare this blazar heating model with recent observational data.

Determining the thermal state of the IGM with observations is challenging, especially for the regions around or below the mean density. The IGM is mostly observed through absorption lines in the spectra of distant quasars,  due to a tiny fraction of neutral hydrogen \citep{1971ApJ...164L..73L}. The so-called Ly$\alpha$ forest can be observed with ground-based facilities for $z\geq 1.7$ while the \textit{Hubble Space Telescope} is currently the only available facility for low redshift ($z\lesssim 0.5$) measurements. As such, we can only observe sections of the  thermal history of the IGM.

A variety of statistics have been developed to analyse spectra from a wide range of instruments. However, deriving the physical parameters of the IGM from the different observables requires a careful calibration to cosmological simulations \citep[see e.g.][]{1997ApJ...489....7R,2000MNRAS.318..817S,2013MNRAS.436.1023B,Bolton_17_sherwood,Gaikwad_2020}.
 
The derivation of the temperature  is also very sensitive to the assumed reionisation and heating history due to a (partial) degeneracy between pressure smoothing and instantaneous temperature. 

Because of the intrinsic observational challenges, numerical shortcomings and difficulty to establish the validity of the different statistics, as well as their sensitivity to different overdensities, accurate constraints on the IGM thermal state are difficult to obtain. While a consensus has arisen about the detection of heating due to HeII reionisation in mildly overdense gas \citep[e.g.][]{2013MNRAS.436.1023B, Gaikwad_2020}, there is no observational consensus yet on the thermal state of the very low-density IGM in which the
impact of blazar heating should be strongest.

Different observational diagnostics suggest or disfavor the presence of blazar heating in the low density IGM. Based on the probability distribution function (PDF) of the transmitted Ly$\alpha$ flux, \citet{2007MNRAS.382.1657K,2008MNRAS.386.1131B,2009MNRAS.399L..39V,2012MNRAS.422.3019C} find that a flat or inverted  $T-\rho$ relation is in good agreement with the data. However, the flux PDF can be strongly affected by systematic errors in the continuum placement \citep{2012ApJ...753..136L} and sample variance \citep{2013MNRAS.428..540R}. Analysis of the curvature of the Ly$\alpha$ spectrum \citep{2011MNRAS.410.1096B,2014MNRAS.441.1916B} prefers a warmer temperature of the IGM. Unfortunately this method, which is based on the smoothness of the spectrum, is mostly sensitive to densities above the mean, especially at low redshift. Finally, Voigt profile fitting of the spectra yields the distribution of line-width ($b$) versus HI column density ($N_\rmn{HI}$). Using the lower envelope of the distribution as a proxy for the $T-\rho$ relation, \citet{Rudie_2012_T_rho,2014MNRAS.438.2499B,Bolton_17_sherwood} find no evidence of blazar heating, but are mostly sensitive to mildly ovderdense ($\Delta\gtrsim1$) gas \citep[also see the discussion in][]{2018MNRAS.474.2871R}.

A strong case for blazar heating comes from a unique spectrum with high signal-to-noise  (SNR) \citep{2017MNRAS.466.2690R}.  The high SNR allows the authors to rescale the optical depth to enhance the signal from low-density regions. At the considered redshifts ($2.5\leq z \leq 3$), they do find that the 
high end of the transmitted flux probability distribution is better matched by a broken powerlaw for the temperature-density distribution, with an inverted slope at the low density end. Their model including temperature fluctuations at low densities produces a satisfactory match as well. The authors also perform a careful analysis of the power spectrum and lower envelope of the ``$b-N_\rmn{HI}$'' distribution and show they are largely  insensitive to the low-density regions.

In this paper, we compare our model for inhomogeneous blazar heating with observational data. Our work builds on the results of \citet{Puchwein_12_Lya} and includes more recent observational data.  We start with a reminder of the main properties of inhomogeneous blazar heating (\S\ref{sec:biased}). We then describe the numerical simulations we use to model the IGM (\S\ref{sec:sims}) and the resulting observables we derive (\S\ref{sec:obs}). We then discuss the direct comparison with observations, suggest a unifying interpretation of Ly$\alpha$ and gamma-ray data (\S\ref{sec:discussion}), and conclude (\S\ref{sec:conclusion}).

\section{Main properties of inhomogeneous blazar heating}\label{sec:biased}

Here we recall the main characteristics of the temperature-density distribution of the IGM under different heating assumptions, such as modelling or ignoring the spatial fluctuations in the blazar heating rate. We refer the reader to Paper I for a more detailed description. Following \citet{2012ApJ...752...23C}, we model the blazar uniform heating rate per comoving volume at a given redshift $\dot{Q}$ by

\begin{eqnarray}
  &  \log_{\mathrm{10}} \left(\frac{\dot{Q}(z)}{\dot{Q}(z=0)}\right)  =0.0315\times[(1+z)^3 -1] \label{eq:redshift_evolution}\\ 
  &-0.512\times[(1+z)^2-1]+2.27 \times [(1+z)-1]. \nonumber
\end{eqnarray}

As TeV blazar abundances at high redshift are observationally poorly constrained, we assume that the blazar luminosity density has a similar redshift evolution as the observed quasar luminosity density. The redshift dependence of Eq.~(\ref{eq:redshift_evolution}) has been chosen to be consistent with the quasar luminosity functions of \citet{2007ApJ...654..731H}. The $z=0$ normalization of the TeV luminosity density can be obtained from \textit{Fermi} observations of the local TeV blazar population \citep[see][]{2012ApJ...752...22B}.
Here we specifically focus on the \textit{intermediate} heating model of \citet{Puchwein_12_Lya} which assumes $\dot{Q}(z=0)=1.08 \times 10^{-7}$ eV Gyr$^{-1}$ cm$^{-3}$. The homogeneous blazar heating model injects this energy with a spatially constant heating rate per unit volume. Our inhomogeneous blazar heating model has the same total amount of energy injected by blazars but accounts for regions receiving more or less heating according to their proximity to heating sources. 
This is modelled based on an analytic formalism relating the distribution of the heating rate to the underlying dark matter distribution. It results in the  filtering function which  removes small scale fluctuations and enhances fluctuations beyond $\simeq 10\, h^{-1}$ Mpc at $z=4$ and $\simeq 40\,h^{-1}$ Mpc at $z=2$.  The shape and redshift evolution of the  window function is set by the mean free path of the gamma rays combined with the bias of the heating sources. Roughly speaking, the heating rate is then given by the assumed blazar luminosity density (itself proportional to the matter density times a bias factor) convolved with a kernel that accounts for the $\propto r^{-2}$ dilution of the gamma ray flux as well as for the absorption with the appropriate mean free path (see Paper I for full details). While we presented two models in Paper I, with galaxy bias and quasar bias, in this work we focus on the model with quasar bias, which is probably more representative of the bias of TeV blazars and may even be a too conservative assumption (see our discussion in \S\ref{sec:discussion}). 

\begin{figure*}
\centering
\includegraphics[width = .23\textwidth ]{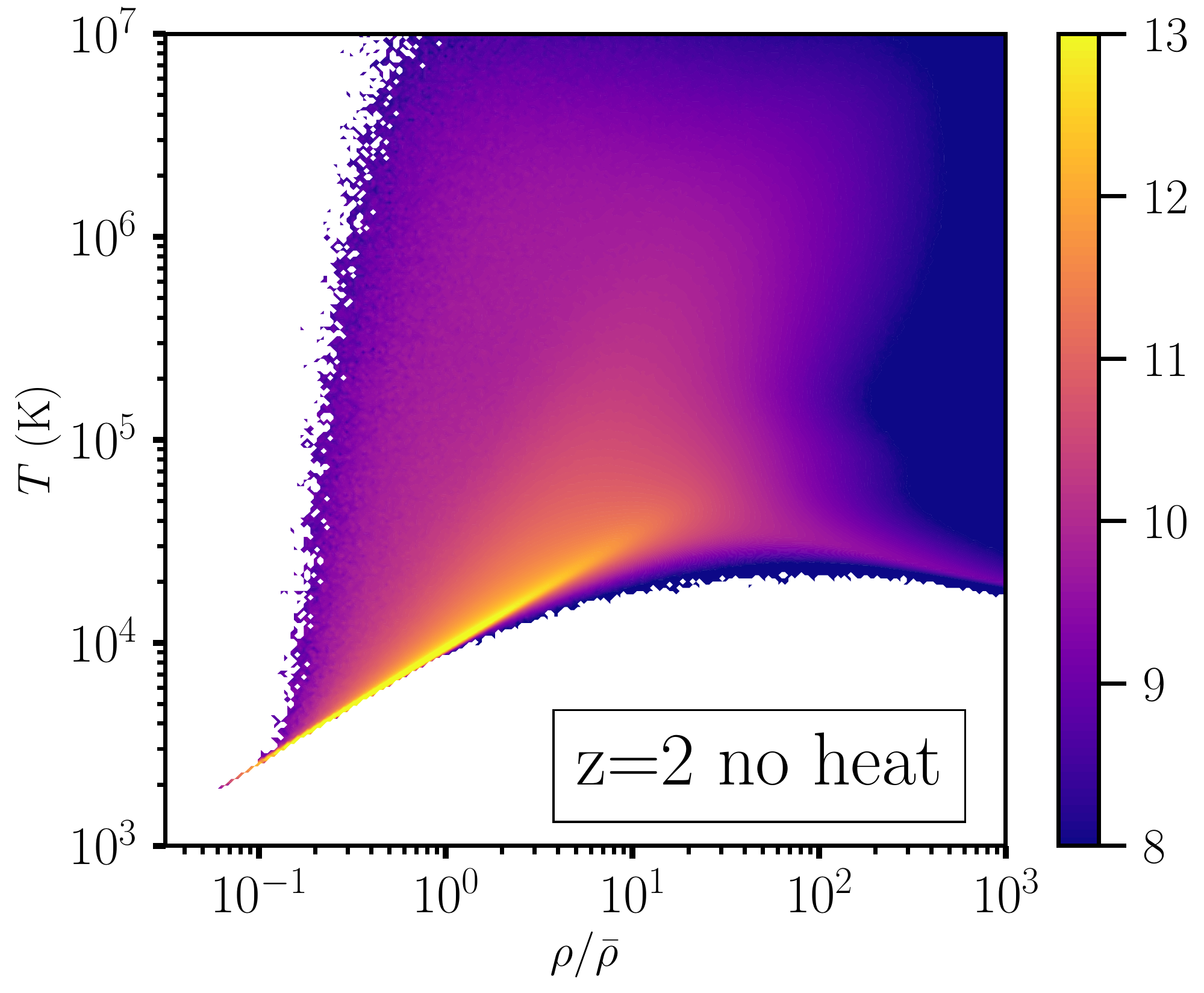}
\includegraphics[width = .23\textwidth ]{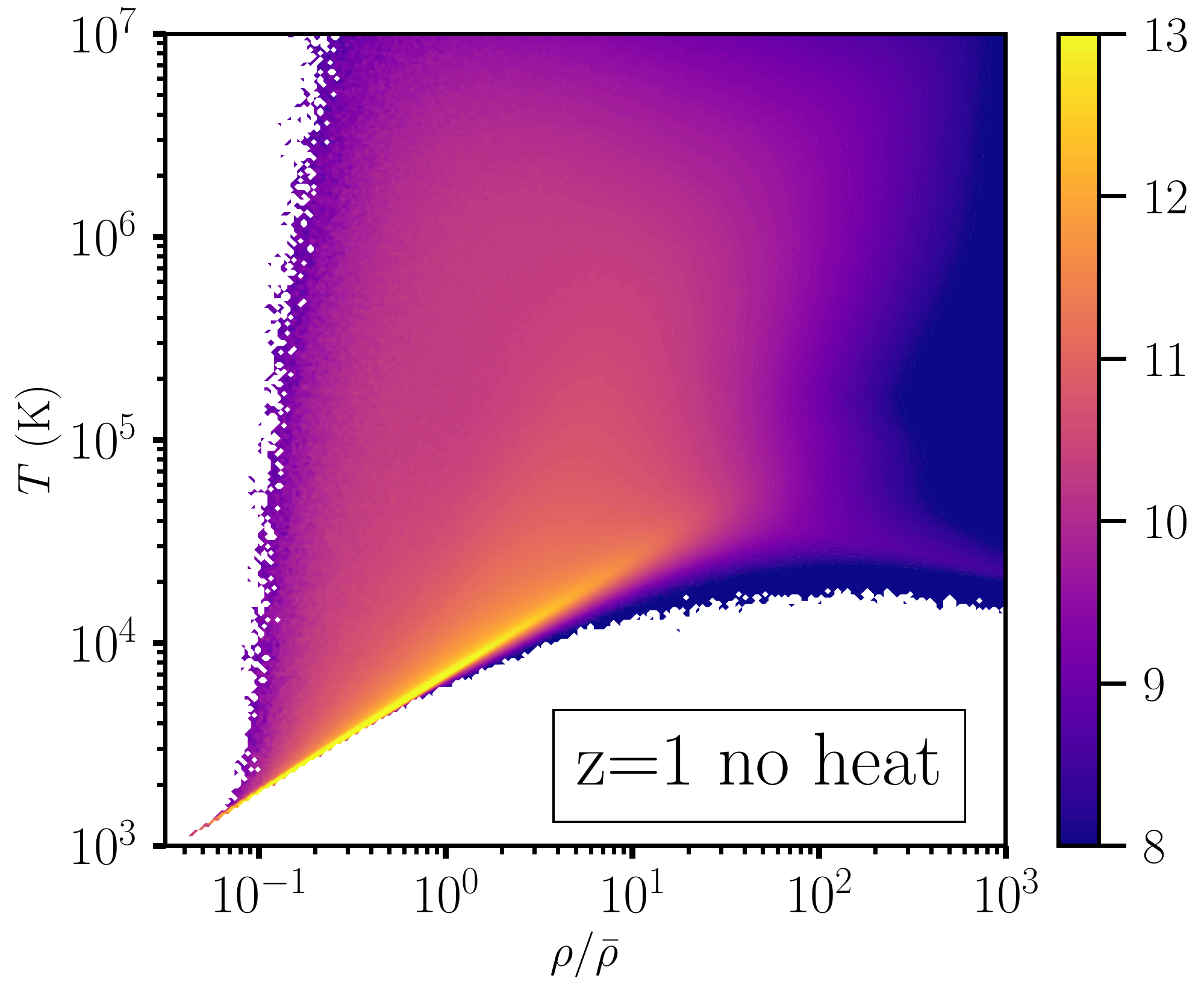}  
\includegraphics[width = .23\textwidth ]{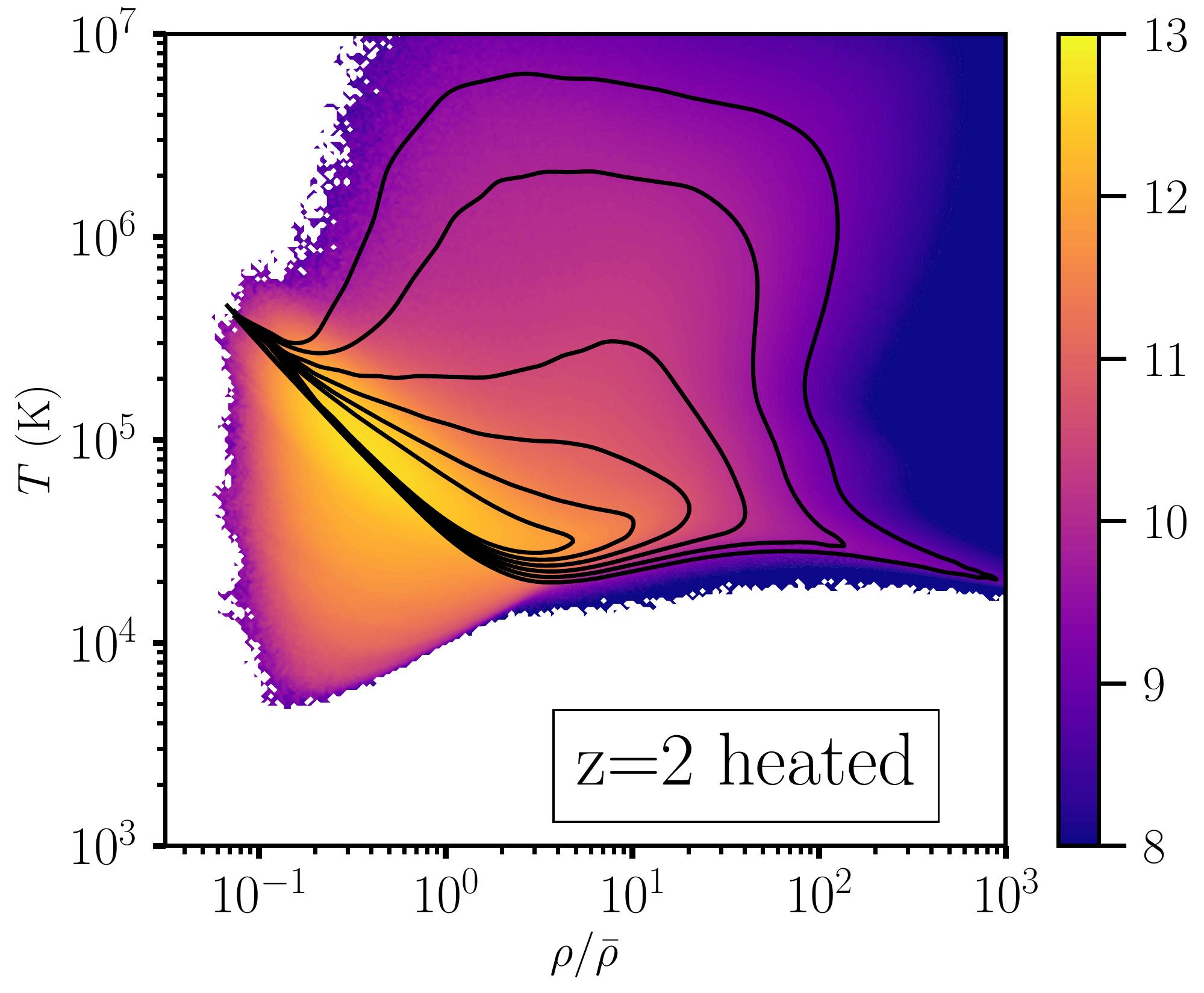}
\includegraphics[width = .23\textwidth ]{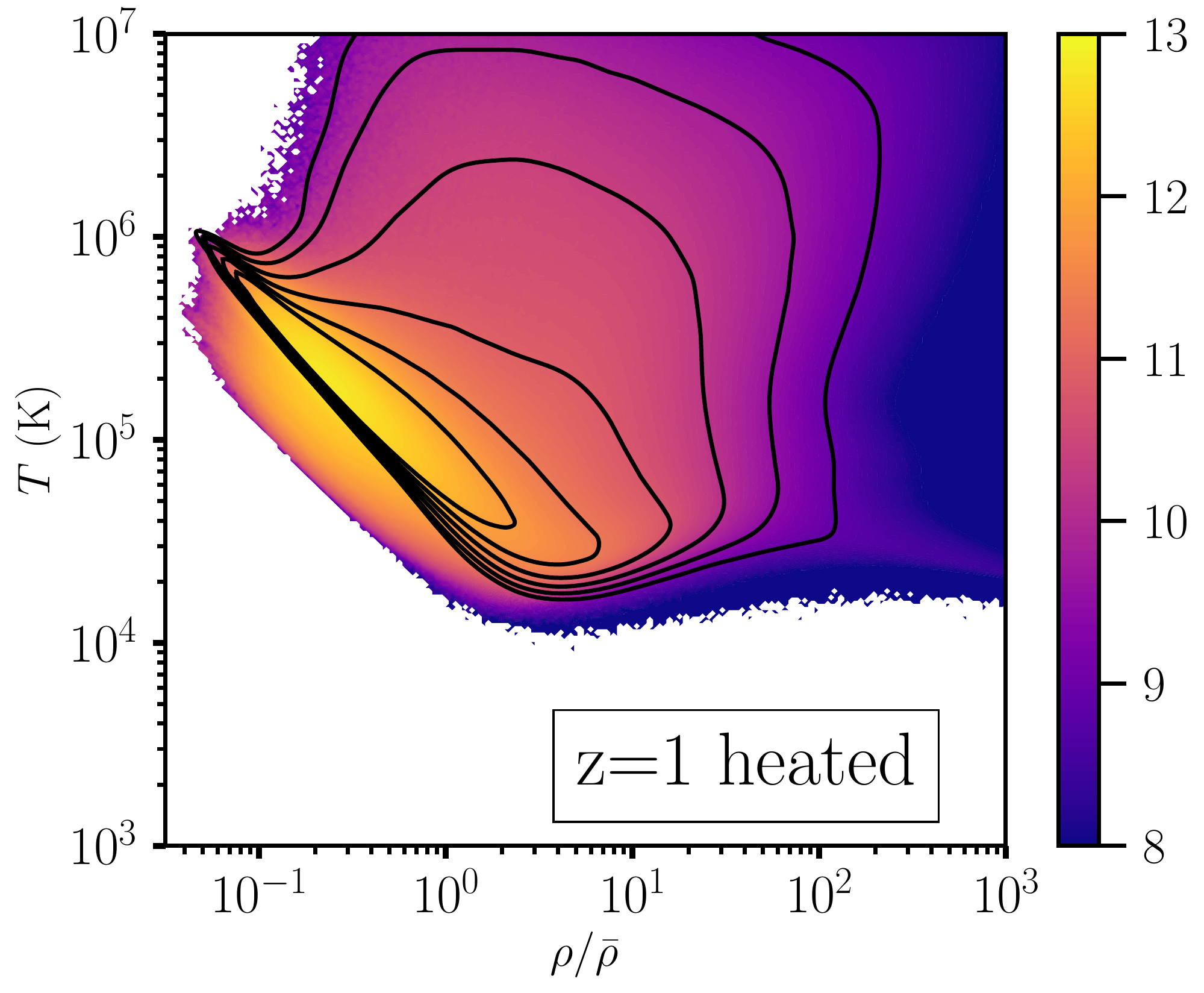}
\caption{ Volume-weighted temperature - density relation at redshifts  $z=2$ and $1$ for the simulations with no blazar heating (two left-hand panels) and with inhomogeneous blazar heating (two right-hand panels). The overlying black contours show the corresponding $T-\rho$ relation for uniform blazar heating \citep{Puchwein_12_Lya} for the same redshift range. The color scale is logarithmic.}
\label{fig:rho_T}
\end{figure*}

Figure~\ref{fig:rho_T} shows the temperature-density distribution according to our inhomogeneous heating model  (two right-hand panels) and when blazar heating is absent (two left-hand panels). As time goes by, the cumulative impact of blazar heating increases, with a higher temperature difference with respect to the unheated case. Still, as some regions are too far from heating sources, they remain cold, as can be seen by the remnant cold gas, especially at $z=2$. This is very different from the uniform heating model (shown by the black contours) and can potentially reconcile the blazar heating model with observations of absorption lines with Doppler parameter ($b\leq 20$ km s$^{-1}$). At the redshifts most accessible with the Ly$\alpha$ forest, the temperature range  covers almost two orders of magnitude at the lowest density and  even around mean density, there is an important spread. At lower redshifts, the whole IGM gets heated up and  for $z\leqslant1$, the inhomogeneous model recovers the uniform model.

\begin{figure*}
\centering
\includegraphics[width = .3\textwidth ]{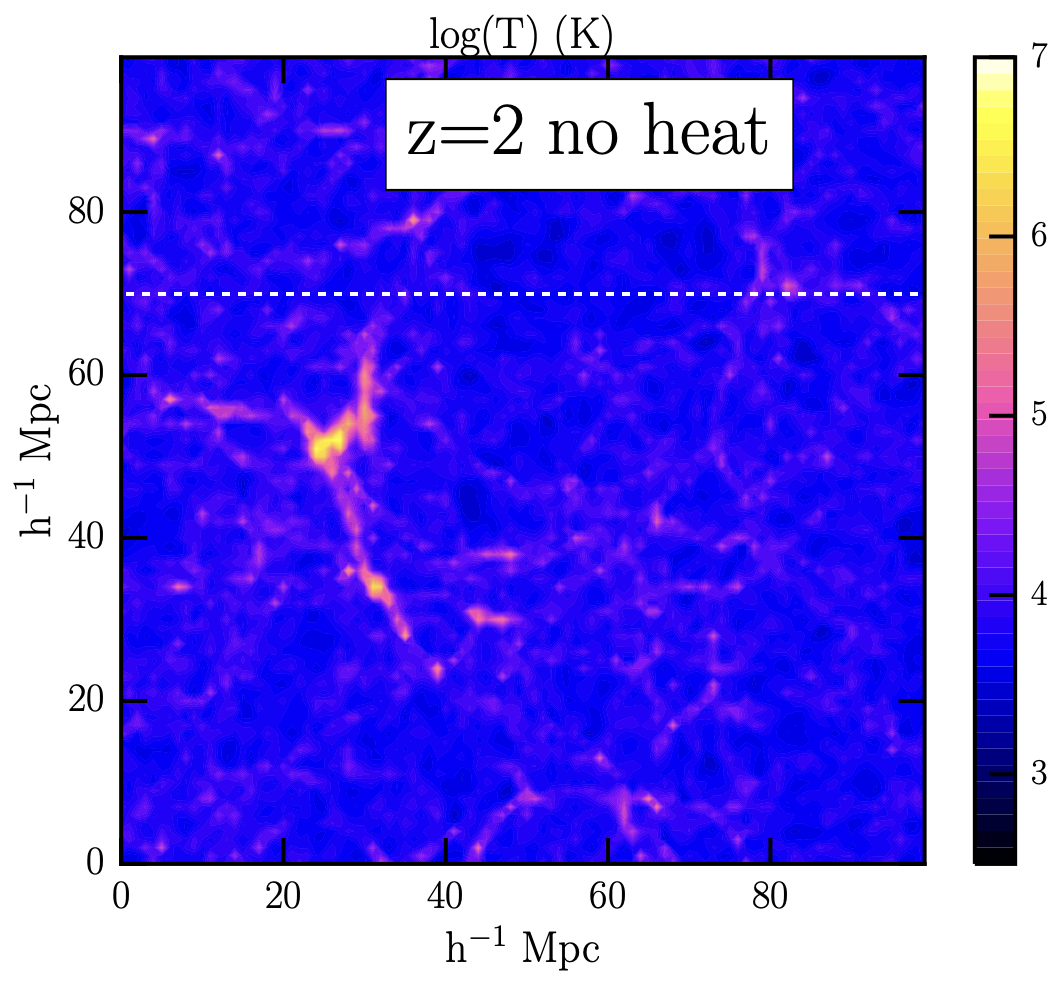}
\includegraphics[width = .3\textwidth ]{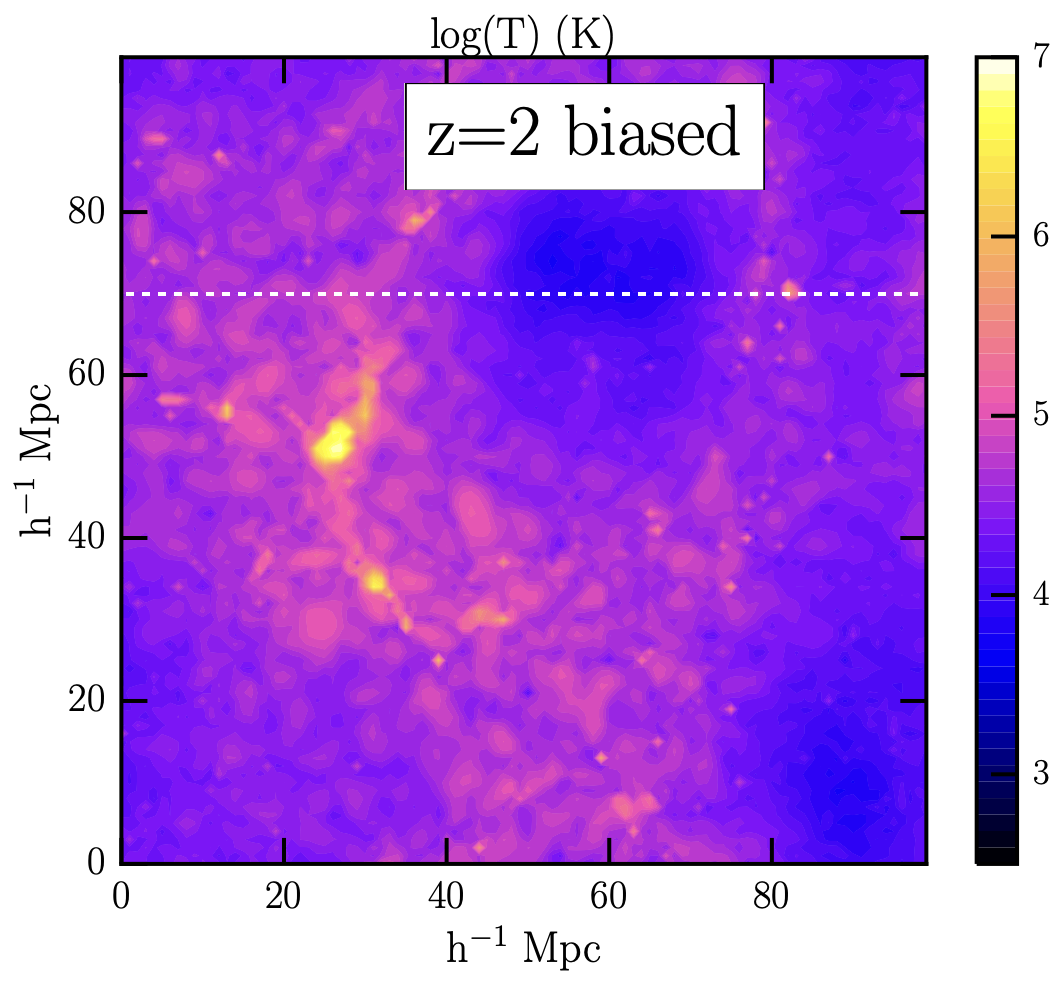}
\includegraphics[width = .3\textwidth ]{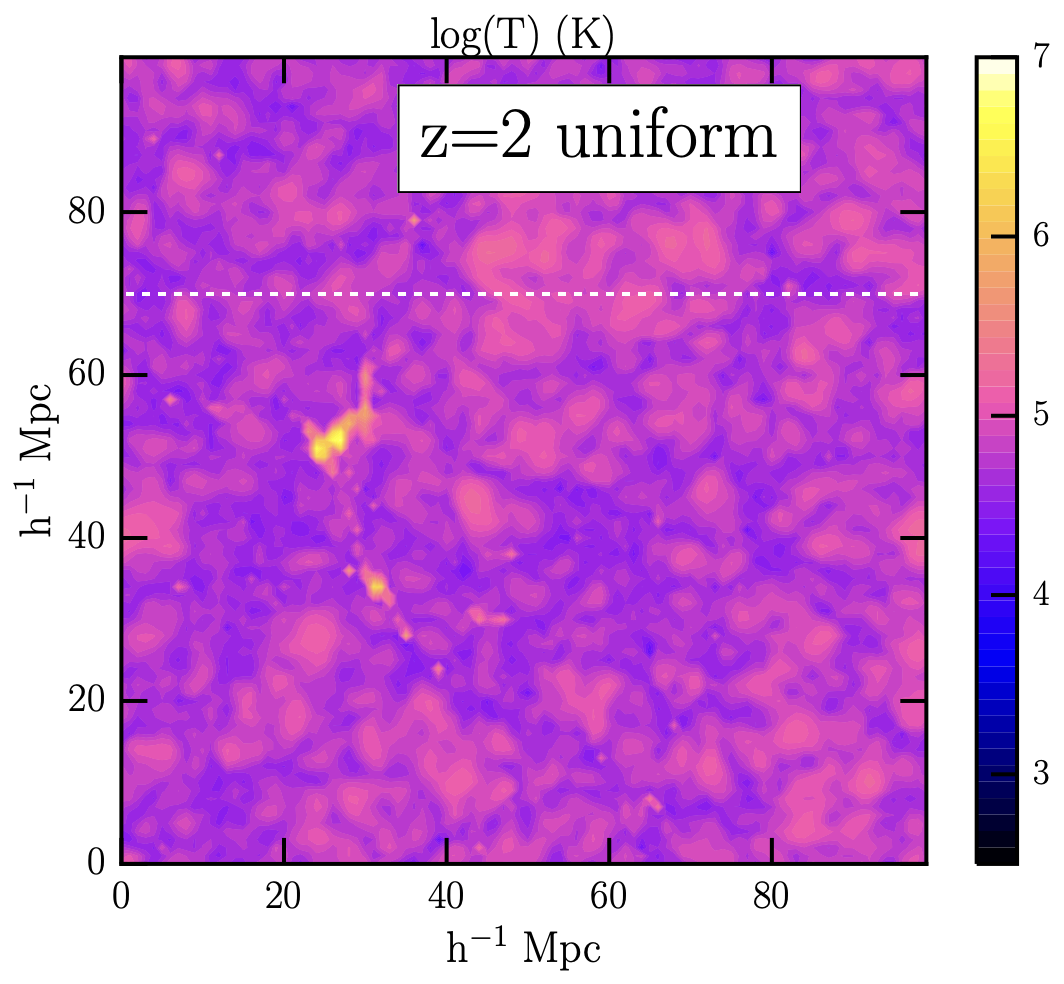}
\caption{Distribution of the temperature in the midplane of the simulation domain at $z=2$ for the unheated case (left), inhomogeneous heating case (middle) and the uniform heating model (right). The white dashed line shows a line of sight that we illustrate in more detail in Fig.~\ref{fig:los}}.
\label{fig:T_flucs}
\end{figure*}

The physical size of the temperature fluctuations in the IGM is mostly set by the mean free path of TeV photons, and is a few tens of Mpc at $z=3$ and increases up to $\simeq$  Gpc in the present day universe. Fig.~\ref{fig:T_flucs} shows a slice through the midplane of our simulation at $z=2$ for the three heating models we considered.  The regions influenced by blazar heating are typically tens of Mpc across. As such, our computational volume should be as large as possible to sample several regions in different thermal states. In the next section, we describe the simulations we perform to model the IGM and how we postprocess them to extract mock observables. In section \S\ref{sec:obs} we show how the thermal properties shown in Figs.~\ref{fig:rho_T} and ~\ref{fig:T_flucs} translate into observables.

\section{Methods}\label{sec:sims}

\subsection{Cosmological simulations}
Our simulations are very similar to the simulations presented in Paper I and are performed at higher resolution. We perform our simulations with \textsc{GADGET-3}, which is an upgraded version of \textsc{GADGET-2} \citep{2005MNRAS.364.1105S}.  It is based on a Smoothed Particle Hydrodynamics (SPH) scheme and solves the gravitational evolution of gas and dark matter with a TREE-PM $N$-body method.  The equations of hydrodynamics are solved with the entropy conserving scheme of \citet{2002MNRAS.333..649S}.   Our simulations use the cosmological parameters inferred from the \textit{Planck} data combined with lensing, \textit{WMAP} and high multipole measurements \citep{2014A&A...571A..16P}: $\Omega_M = 0.305$, $\Omega_{\Lambda} = 0.694$, $\Omega_B = 0.0481$, $h = 0.679$, $\sigma_8 = 0.827$ and $n_s$ = 0.962.  These values are slightly updated with respect to the simulations presented in \citet{Puchwein_12_Lya} and \citet{2015ApJ...811...19L} but we do not expect any significant impact on the results presented here. In all cases, our simulations start with the same initial conditions at $z=110$. The simulation domain has a comoving side length of 100 $h^{-1}$ Mpc and periodic boundary conditions. We use $N= 2\times 1280^3$ particles, which yields a mass resolution of $m_\mathrm{gas}=1.5\times10^{6} h^{-1} M_{\odot}$ and $m_\mathrm{DM}=7.2\times 10^{6} h^{-1} M_{\odot}$ for baryonic and dark matter particles, respectively. We used a comoving gravitational softening length of 3.9 $h^{-1}$ kpc. The size of the box  was chosen to cover the typical length scales of heating fluctuations, of tens of Mpc (see Fig.~\ref{fig:T_flucs}). We perform a set of simulations focusing  on $z\geq 1.75$ for direct comparison with ground-based Ly$\alpha$ data.

\subsection{Modelling the thermal state of the IGM}
As this work focuses on the IGM, we use a simplified model for star formation in our simulations, in which gas particles with  density $\delta_{\mathrm{gas}}\equiv\rho_{\mathrm{gas}}/\bar\rho_{\mathrm{baryon}}-1\geq 1000$ and T $\leq 10^5$ K are directly converted into stars \citep{2004MNRAS.354..684V}. Although this yields inaccurate galaxy properties, it does not affect the low-density IGM and significantly speeds up the simulations \citep{Bolton_17_sherwood}. 
Photoionisation and photoheating is based on a \citet{2012ApJ...746..125H} UV background and the assumption of ionization equilibrium. The simulations in Paper~I and \citet{Puchwein_12_Lya} were based on the \citet{2009ApJ...703.1416F} UV background, which in the absence of blazar heating overpredicts the effective optical depth in the Lyman-$\alpha$ forest at all redshifts. However, the optical depths are rescaled during postprocessing to match the mean transmitted flux from observations by \citet{Becker_13_tau}. This rescaling aims to correct for uncertainties in the assumed photoionisation rate and strongly reduces differences when comparing the observables resulting from simulations with different UV backgrounds. The main residual effect is that slight changes in the photoheating result in somewhat different IGM thermal histories, which in turn affect the pressure smoothing in the simulation and the thermal broadening of predicted absorption lines.

\begin{figure}
\centering
\includegraphics[width=\columnwidth ]{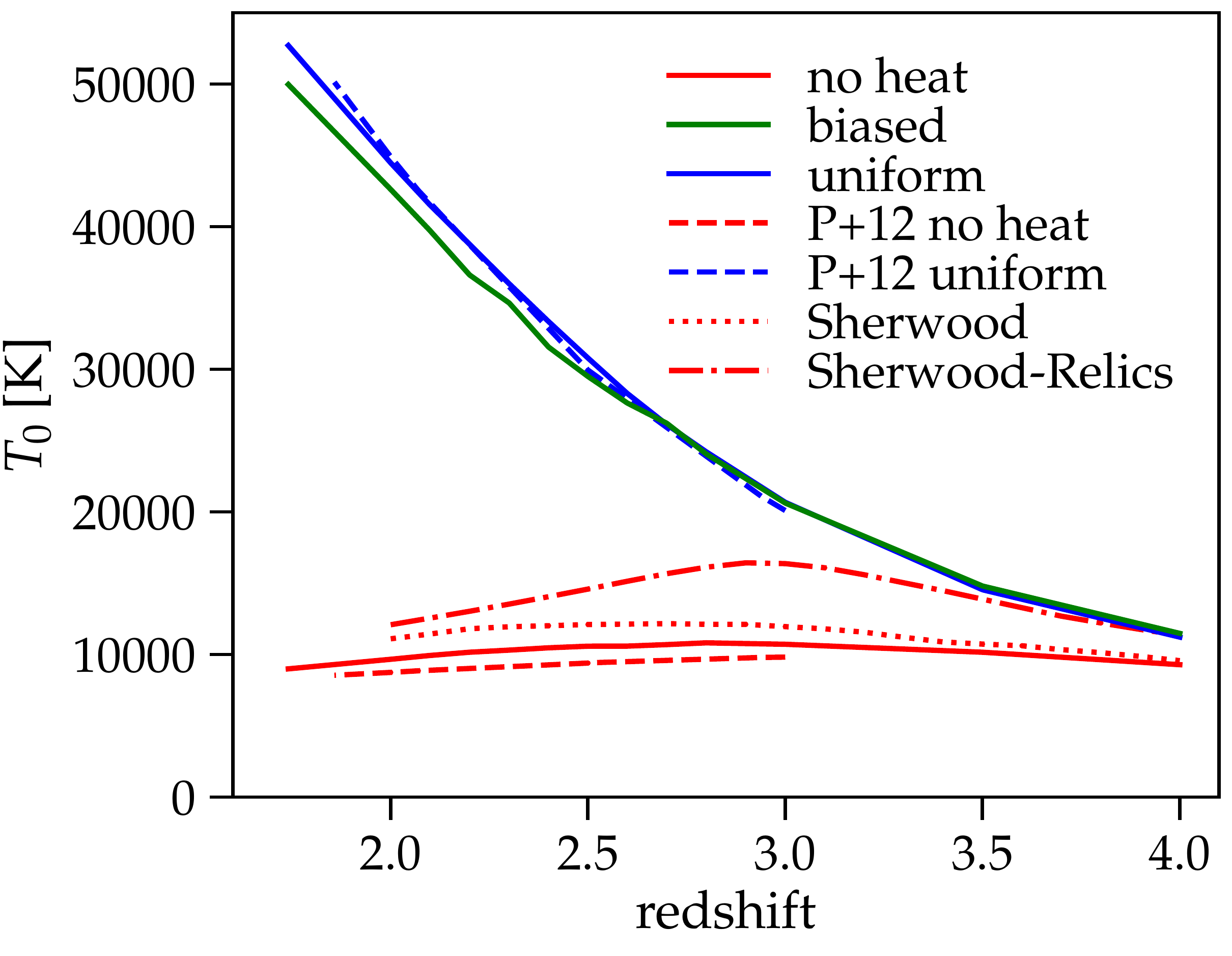}
\caption{Median IGM temperature of gas at mean cosmic baryon density ($0.975<\Delta<1.025$) as a function of redshift in our simulations (solid lines). For comparison results from a previous study investigating blazar heating of the IGM are shown \citep[][dashed lines]{Puchwein_12_Lya}. Also indicated is the thermal evolution of the IGM in the Sherwood and Sherwood-relics simulations, which do not consider blazar heating but account for a larger amount of heating from HeII reionisation.}
\label{fig:T0}
\end{figure}

In Fig.~\ref{fig:T0}, we compare the thermal evolution of the IGM (measured at mean cosmic baryon density) in our different simulations, as well as to results from the literature. We here perform three simulations: one without any blazar heating, one with uniform blazar heating and one with inhomogeneous blazar heating. Due to the different choice of UV background our simulation without blazar heating is slightly hotter than the run without blazar heating in \citet{Puchwein_12_Lya}. The difference is, however, very small compared to the (potential) impact of blazar heating. The blazar heating in our uniform heating model is identical to that in the ``intermediate heating'' model presented in \citet{Puchwein_12_Lya}. The blazar heating results in much higher IGM temperatures at low redshift. Note however that the relative importance of blazar heating also depends on density.

Finally, we show two simulations from the literature in Fig.~\ref{fig:T0}, the 40-2048 simulation from the Sherwood simulation suite \citep{Bolton_17_sherwood} and a simulation of the same volume from the Sherwood-Relics project (Puchwein et al. 2022, in prep.;  also see \citealt{2020MNRAS.494.5091G,2019MNRAS.485...47P}). These simulations do not include heating by blazars but account for a larger (and likely more realistic) amount of heat injection by He~\textsc{ii} reionisation. These runs will be discussed further in Sec.~\ref{sec:discussion}.

\subsection{Producing mock Ly$\alpha$ spectra}
The main output of our simulations are the mock Ly$\alpha$ spectra throughout the entire simulation volume. To obtain these, we extract 100 lines of sight from each output. The lines of sight are randomly selected and aligned with the main axes of the simulation. We use the same lines of sight for all the different heating models, but change the random selection from one redshift to the other. For each line of sight, we determine the density, velocity, temperature and resulting optical depth of HI along 2048 equally spaced bins.  The resulting dataset is the basis of our subsequent analysis.

\begin{figure}
\centering
\includegraphics[width=0.9\columnwidth ]{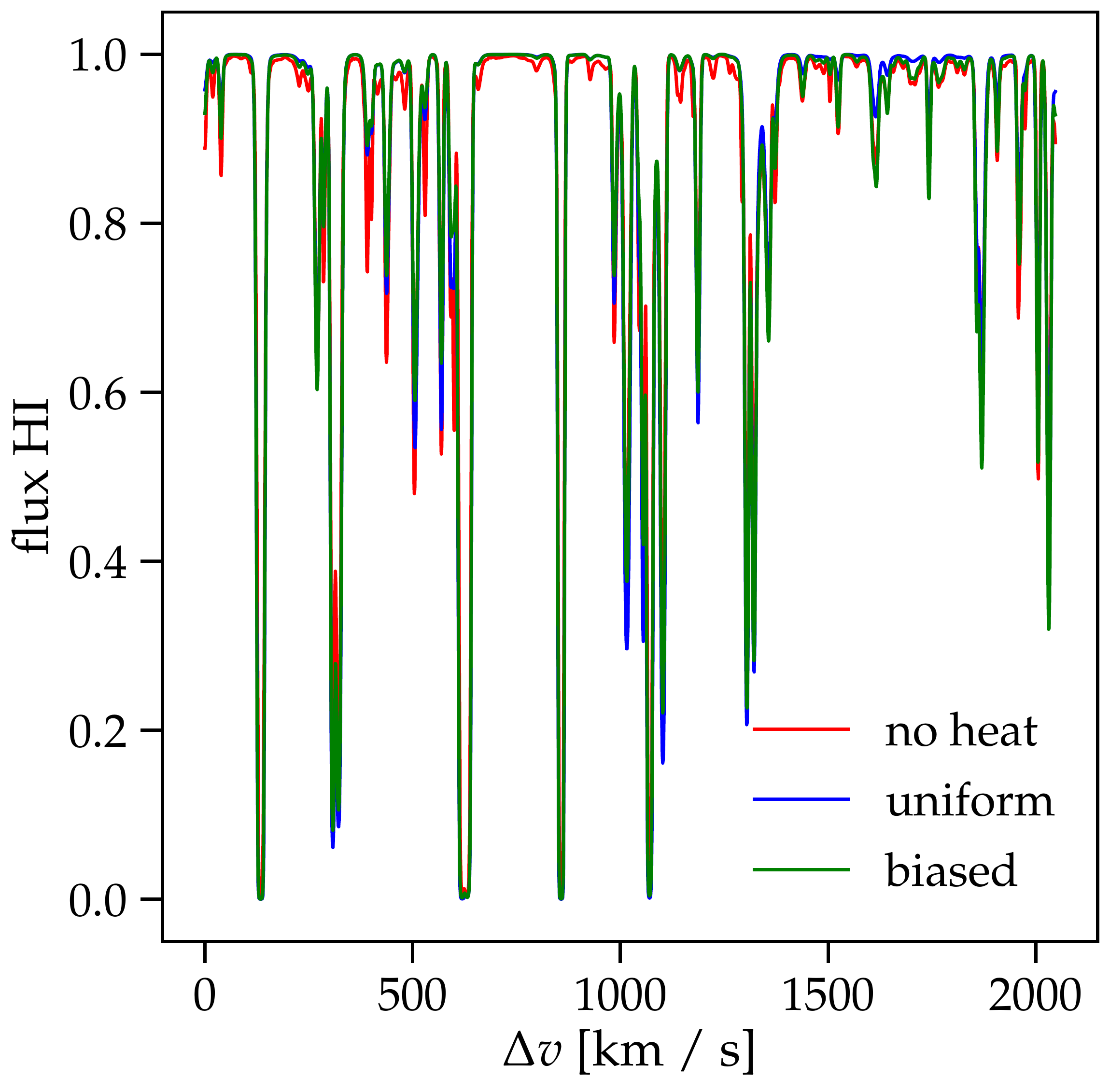}
\caption{Illustration of the transmitted HI Ly$\alpha$ flux at $z=2$ in the different heating models along a line of sight which includes both warm and cold regions. The location of the line of sight is shown on Fig.~\ref{fig:T_flucs}, and is chosen in the midplane of the domain. The optical depths have been rescaled to the mean transmission at the considered redshift.}
\label{fig:los}
\end{figure}

An example is shown in Fig.~\ref{fig:los}, which displays the normalized transmitted HI Ly$\alpha$ flux along one line of sight chosen to illustrate the differences in our heating models. This specific line of sight first crosses a region which is heated in both blazar-heated models (first half of the spectrum, see Fig.~\ref{fig:T_flucs} for the associated temperature map) and then a region which is heated in the uniform model but remains cold in the biased heating model (second half of the spectrum). This effect is clearly visible in the spectrum, where the Ly$\alpha$ flux has an intermediate value between both extreme models.

In order to allow direct comparison of our synthetic spectra with observed data, we rebin the dataset to the spectral resolutions of the different spectrographs and add noise with the same properties as the observations. The outputs of our high-redshift simulations are separated by  $\Delta z=0.1$, starting at $z=3$. This allows us to cover the complete pathlength of a photon from $z=3$ down to $z=1.8$.  For each output, we compute a comoving line-of-sight length consistent with the line-of-sight length covered by the data (Fig. 1 in ~\citet{Rudie_13_CGM}).

Details of the exact computation are provided in the relevant comparisons with observations. In the next section, we show how the shape of the temperature-density distribution and the size of the temperature fluctuations affect the observable properties of the Ly$\alpha$ forest and compare with the observational data.

\section{Comparison with observations}\label{sec:obs}

We compare our simulations with different heating models with observations of the Ly$\alpha$ forest. In this section,  we focus on statistics that are sensitive to low density regions and new observational data. In the Appendix, we provide comparisons with the simulations presented in \citet{Puchwein_12_Lya} for  reference. 

\subsection{Rescaled flux PDF}
\citet{2017MNRAS.466.2690R} recently proposed a new method to probe the thermal state of the low-density IGM.  In this method the optical depth is rescaled by a factor $A$. As a result the flux probability distribution function (PDF) is more focused on the high-transmission regions, which are likely corresponding to the low-density regions. Additionally, \citet{2017MNRAS.466.2690R} rescale the transmitted flux to the value of the 95$^\rmn{th}$ percentile of the distribution, thus reducing the impact of the error in the continuum placement. Their comparison with simulations shows a better agreement with models with additional heating and/or inhomgeneous heating. To compare with the rescaled flux PDF by \citet{2017MNRAS.466.2690R}, we perform the exact same steps as in their analysis.

As described in section 3 of \citet{2017MNRAS.466.2690R}, we first rescale the optical depth so that the mean transmitted flux $F$ matches the observed value ($\langle F_{\mathrm{obs}}\rangle=0.371$) at the central redshift of the observation ($\langle z\rangle=2.75$), using data from \citet{2013MNRAS.436.1023B}. We perform this operation at once for all lines of sight resulting from the same simulation output. Then, as described in their section 4.1, we smooth the flux with a Gaussian of full width at half maximum of $7.2$ km s$^{-1}$ (to model the the line spread function) and rebin it into bins of $\Delta \varv=2.5$ km s$^{-1}$ the pixel resolution of the UVES spectrograph). We then add Gaussian noise with $\sigma=(\sigma_0^2+F(\sigma_\rmn{c}^2-\sigma_0^2))^{1/2}$ with $\sigma_0=0.0028$ and $\sigma_\rmn{c}=0.0088$ (see Appendix F in \citet{2017MNRAS.466.2690R}). We then rescale the transmitted flux to the transmitted flux in the 95$^\rmn{th}$ percentile. The latter is close to the peak of the flux PDF, and is therefore less noisy than the mean. This operation is performed independently for each line of sight, in chunks of size 10$\,h^{-1}$~Mpc. Finally, to enhance the impact of low density regions, we rescale the optical depth by a factor $A=10$, which yields a rescaled transmitted flux $F_A=F^A$.

To reconstruct the pathlength of the Ly$\alpha$ absorbers in the  observed line-of-sight, we stitch together subsections of lines of sight from all outputs between $z=3$ and $z=2.5$. For each output, the size of the subsection is set by the comoving distance $\Delta l$ corresponding to $\Delta z=0.1$ for the considered redshift. The first pixel of the subsection is randomly chosen along one  of the lines of sight and we use the periodic boundary conditions to complete the line of sight if the edge of the computational box is reached before $\Delta l$ is completed. The lines of sight are randomly chosen for each output. While this produces small discontinuities at the junctions of the subsections, it is a more realistic representation of the varying cosmic structure that would be encountered by a photon \citep[see e.g.][for a discussion]{Hummels_17_Trident}. This method implicitly assumes that the removal of contaminants in the observed sightline (such as metals) does not affect the redshift distribution of the Ly$\alpha$ absorbers along the path covered by  the quasar sightline.

\begin{figure*}
\centering
\includegraphics[width = .9\textwidth ]{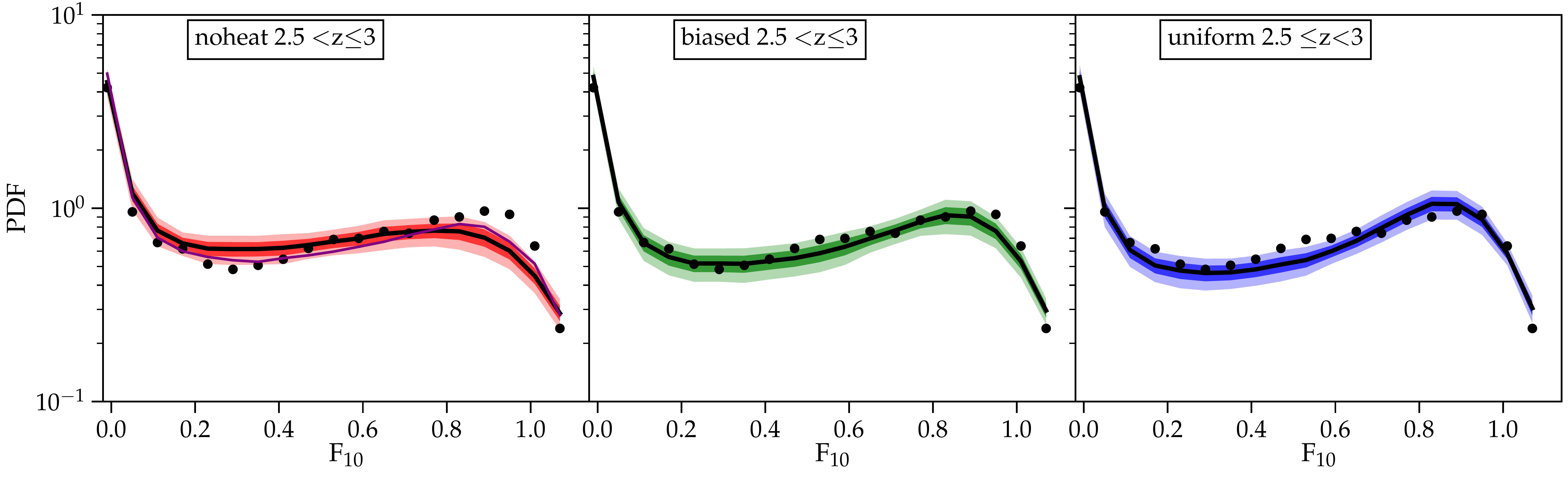}
\caption{Probability distribution function of the rescaled flux in the unheated (left), inhomogeneously heated (middle) and uniformly heated model (right) between $2.5 <z\leq 3$. The thick line represents the mean over the 100 lines of sight, and the dark and light shaded area the 1 and 2 $\sigma$ (standard deviation) uncertainty intervals around the mean (indicating both sample variance and the effects of noise). The observed PDF is given with the black dots \citep{2017MNRAS.466.2690R}. It is worth keeping in mind that it is based on a single line of sight with exquisite data quality. The left panel includes the mean value of the rescaled flux PDF from the Sherwood-Relics simulation (purple line). Following~\citet{2017MNRAS.466.2690R}, no error bars are added to the data points as uncertainties are model dependent and computed separately for each simulation.}
\label{fig:PDF_A}
\end{figure*}

Figure~\ref{fig:PDF_A} shows the PDF of $F_A$ in our three different heating models, compared with the observed PDF.  As the observed line of sight is unique, cosmic variance can significantly affect the resulting flux PDF \citep{2013MNRAS.428..540R,2017MNRAS.466.2690R}. To provide some measurement of the spread of the flux PDF, we recreate a total of one hundred  lines of sight from each of our simulations (following the method described above). The colored shaded areas show the resulting one and two sigma deviations from the mean. While we are confident that we do not oversample specific locations of the simulation, in Paper~I we showed that the typical length scale  of heated/unheated regions is of a few tens of Mpc/h. As such, the number of regions sampled by our simulation is somewhat limited and we likely underestimate cosmic variance. 

Our simulations show that the rescaled flux PDF of the Ly$\alpha$ forest cannot be reproduced by a model without additional heating before or around $2.5<z<3$. Inhomogeneous blazar heating does provide a flux PDF compatible with observations. The uniform model, where the same total amount of blazar heating is injected, does not agree equally well with the data. The model lies beyond 2$\sigma$ of a third of the data points. This model has less physical motivation, as blazar heating is naturally inhomogeneous, because of the biased distribution of the sources. 

For comparison, we also show results for the Sherwood-Relics simulation (purple line, leftmost plot of Fig. \ref{fig:PDF_A}), which does not include blazar heating, but predicts more heating from HeII reionization. The latter is a consequence of using a more accurate non-equilibrium ionization/heating solver and a UVB that produces a realistic reionization history \citep{2019MNRAS.485...47P}. The increased HeII heating brings the rescaled flux PDF in the absence of blazar heating in somewhat better agreement with the data, but not to the extent that our biased blazar heating model does. We will discuss this further in Sec.~\ref{sec:discussion}.

\subsection{Properties of absorption lines}

The statistical properties of the line-width $b$  and column density $N_{\mathrm{HI}}$ of the individual Ly$\alpha$ absorption lines carry information about the density and temperature of the IGM. \citet{Schaye_99_IGM_temp} determined that it can be used as a proxy for the temperature-density relation in the low-density regime by assuming that, for a given column density, the smallest Doppler width is only set by thermal broadening while higher values result from turbulent motions within the absorbers. Intrinsically, this method relies on a unique relation between temperature and density, whereas the inhomogeneous blazar heating model shows a wide distribution (see Fig. \ref{fig:rho_T}).  Although the complete $b-N_{\mathrm{HI}}$ distribution carries information on the physical properties of the IGM, often only the lower envelope of the distribution is considered \citep[but see the recent work by][]{Hiss_19_bNhfull}.

We compare our simulations with the data from \citet{Rudie_12_CGM}, based on 15 quasar absorption lines, as part of the Keck Baryonic Structure Survey \citep[KBSS]{Steidel_10_KBSS}. The sample covers a total pathlength $\Delta z=8.27$ (comoving pathlength $\Delta l=$ 26.9) between $2.02\leq z \leq 2.84$ with a mean redshift $\bar{z}=2.34$. The data was obtained with the HIRES spectrograph, with a FWHM$\simeq 7$ km s$^{-1}$ and typical signal-to-noise ratio ~$\simeq 50-250$. These observations increase the number of high signal-to-noise Ly$\alpha$ absorbers in that redshift range by a factor of 2. Thanks to this large sample size, we can consider the complete 2D distribution of the absorbers in the $b-N_{\mathrm{HI}}$ plane, rather than limiting ourselves to their 1D projections. We specifically focus on low column densities, log$(N_{\mathrm{HI}}) \leq 14$, corresponding to low density optically thin absorbers. Those are the most likely not associated with galaxies, but with the IGM \citep{Rudie_13_CGM}. In this section, we compare our simulations with the \citet{Rudie_12_CGM} data, and follow their analysis as closely as possible.

Before extracting the physical information from our mock spectra, we add mock noise based on the noise distribution from the data.  The average SNR is 100, but varies between 50 and 250, with the highest values at high redshift.  Based on the histograms of the SNR of the individual pixels of each quasar, we determine the redshift-dependent probability  distribution function of the SNR. For each simulated line-of-sight we then randomly choose the SNR from the distribution according to its redshift. Overall, the resulting line properties are, however, in agreement with the line properties from mock data obtained with a simpler, fixed SNR set to 100 (see the third column in Fig.~\ref{fig:systematics}).

In order to determine the line widths $b$ and column densities $N_{\mathrm{HI}}$ of the absorbers, each of them has to be identified and individually fit. \citet{Rudie_12_CGM} use the VPFIT software \citep{Carswell_VPFIT}, which automatically finds the properties ($z$, $b$,  log~$N_{\mathrm{HI}}$) of the absorbers using a $\chi^2$-reducing method. Lines with $b \leq 8$ or $b \geq 100$ or relative errors larger than 50 per cent for the properties of the absorbers are discarded from the initial sample. To enable the closest possible comparison, we also use VPFIT to analyse our lines of sight. Given the size of our simulations, each line of sight contains too many absorbers ($\gtrsim 150$), making the convergence of VPFIT  very slow. Following \citet{Bolton_17_sherwood}, we divide each line of sight into 5 equally-sized chunks and analyse each of them separately. At the end, we concatenate the absorbers from the 5 sections into one list. We remove the absorbers within  $\Delta \varv=500$ km s$^{-1}$ of the section edges. Visual inspection determined that this is a satisfactory way to eliminate boundary effects while maintaining most of the sample. We have tested cuts up to 1000 km s$^{-1}$ and find no significant difference in our results. We perform the same additional cuts as \citet{Rudie_2012_T_rho} and also use their outlier removal procedure aimed at removing narrow metal lines from their sample. Although our mock spectra do not contain metal lines by construction, we perform the same outlier removal procedure to mimic accidental removal of hydrogen lines and find it has limited effect.

To allow for a statistical comparison with the data, we produce 100 mock samples, each with a pathlength distribution  $\Delta l(z)$ based on Fig.~1 in \citet{Rudie_13_CGM}, i.e. similar to the observed sample. To reconstruct each individual mock sample, we appropriately combine randomly selected individual absorbers using a redshift-dependent probability for an absorber to arise per unit path length from our different  output redshifts.

\begin{figure*}
\centering
\includegraphics[width = .95\textwidth ]{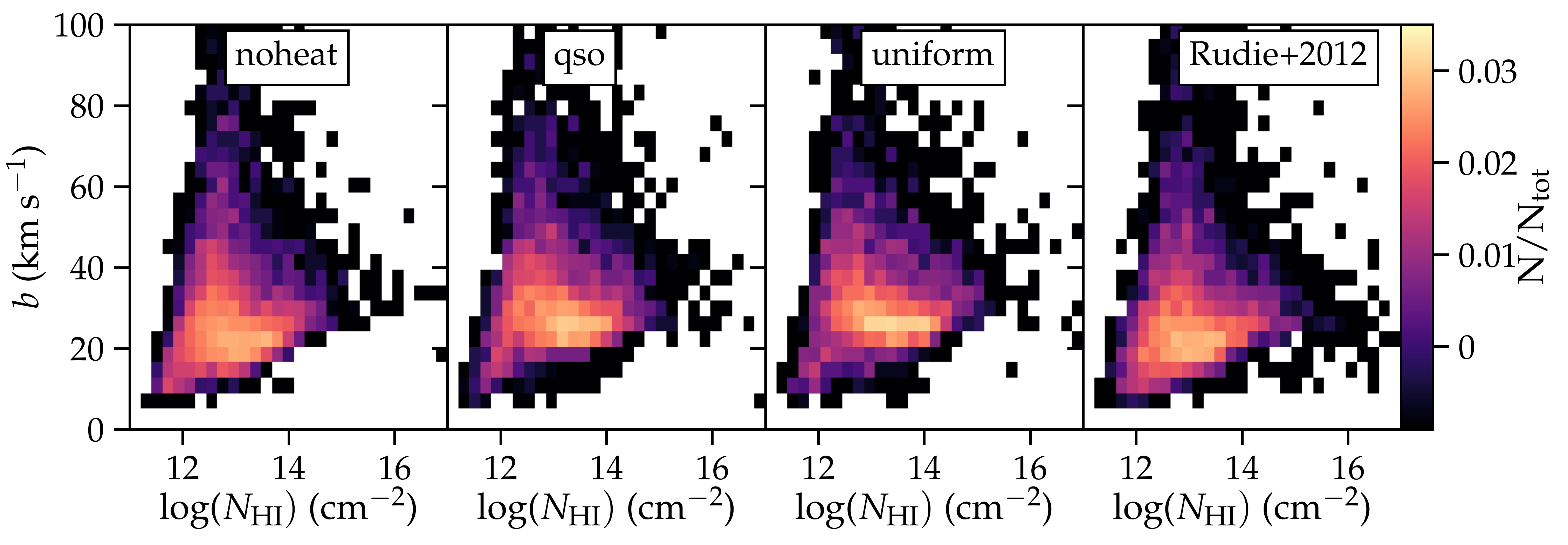}
\caption{Line-width versus column density distribution in the three different heating models, compared with the data from \citet{Rudie_12_CGM}. The redshift range extends from $z=2.02$ to $z=2.84$ with a mean value $\bar{z}=2.34$.}
\label{fig:b_Nh}
\end{figure*}

In Fig.~\ref{fig:b_Nh} we show the complete $b-N_{\mathrm{HI}}$ distribution in the three heating models compared with the data (right panel). From these plots it is clear that the complete 2D distribution is different in the three heating models and that one may loose information by considering only its 1D projections.  The broad temperature-density distribution in the inhomogeneous heating case is reflected in a broader $b-N_{\mathrm{HI}}$ distribution at low column densities, hence representing an intermediate case between the unheated and the uniformly heated case. By eye, the ~\citet{Rudie_12_CGM} sample looks closest to the unheated model. Although the inhomogeneous heating model seems to reproduce the lower envelope of the observed distribution, the peak of both blazar heated distributions show an offset.

\begin{figure*}
\centering
\includegraphics[width = .95\textwidth ]{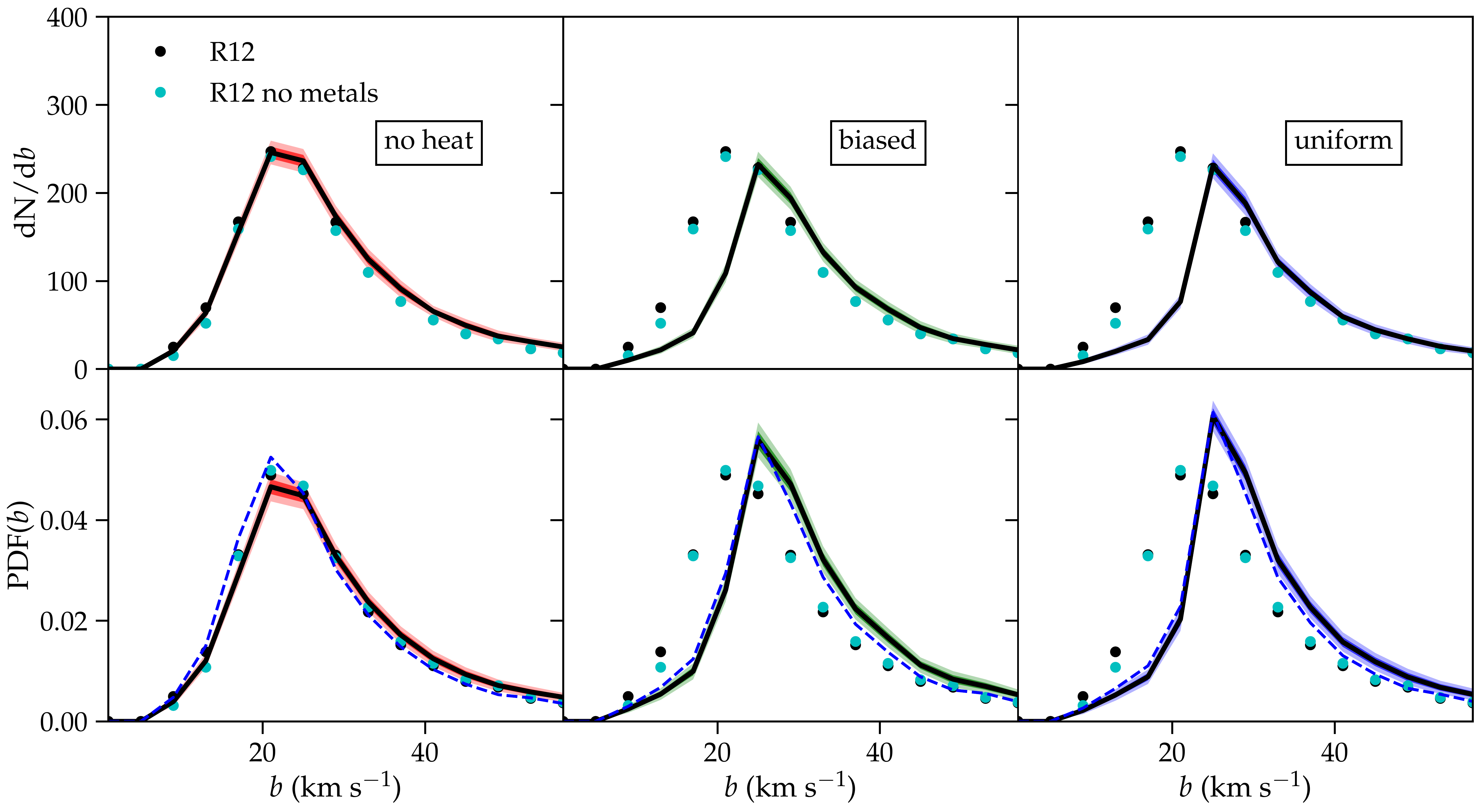}
\caption{Normalised (bottom) and un-normalized (top) line-width distribution in the three different heating models, compared with the data from \citet{Rudie_12_CGM}. The black dots show the complete dataset and the cyan point shows the data resulting from the outlier removal procedure. The colored shaded areas show the one and two sigma confidence regions around the mean.  The blue dashed lines shows the PDF with a resolution correction applied, which was derived from the Sherwood simulation suite (see \S\ref{sec:discussion}). The redshift range extends from $z=2.02$ to $z=2.84$ with a mean value $\bar{z}=2.34$ and the column density range is $[10^{11}-10^{14.5}]$ cm$^{-2}$.}
\label{fig:hist_b}
\end{figure*}

\begin{figure*}
\centering
\includegraphics[width = .95\textwidth ]{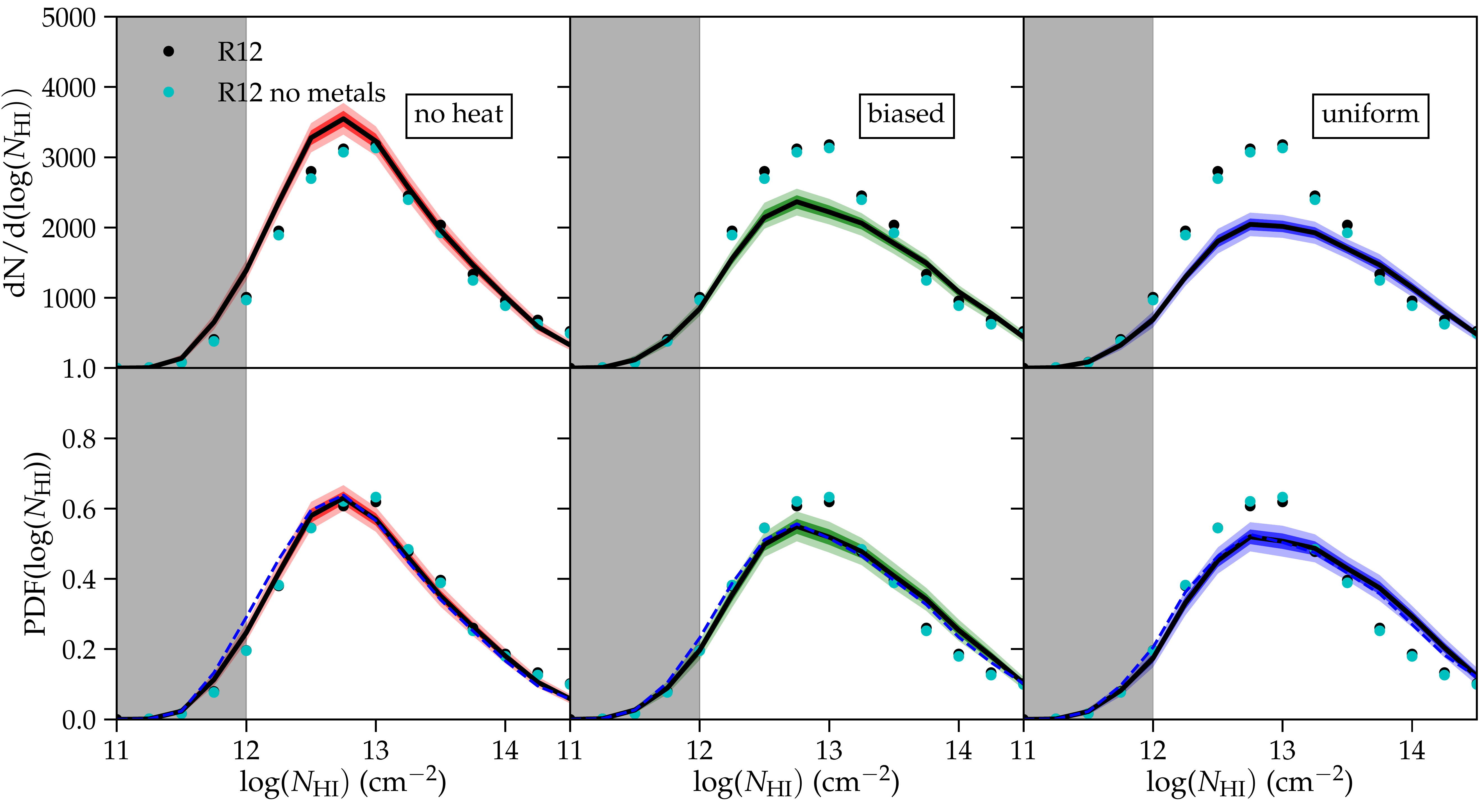}
\caption{Normalised (bottom) and un-normalized (top) PDF of the column density between 10$^{11}$ and 10$^{14.5}$ cm$^{-2}$ in the three different heating models, compared with the data from \citet{Rudie_12_CGM}. The black dots show the complete dataset and  cyan point shows the data resulting from the outlier removal procedure. The shaded areas show the one and two sigma confidence regions around the mean. The grey shaded area indicates regions where the line-fitting algorithms  are less reliable, potentially affecting the results.  The blue dashed lines shows the PDF with a resolution correction applied, which was derived from the Sherwood simulation suite (see \S\ref{sec:discussion}). The redshift range extends from $z=2.02$ to $z=2.84$ with a mean value $\bar{z}=2.34$.}
\label{fig:hist_Nh}
\end{figure*}

Figures~\ref{fig:hist_b} and~\ref{fig:hist_Nh} show a quantitative comparison of the ~\citet{Rudie_13_CGM} data and the heating models. We show the complete dataset from ~\citet{Rudie_13_CGM} as well as the subset where likely metal lines have been removed. The impact of the metal removing procedure on the global line properties is small.  We show the total number of lines as a function of their linewidth (Fig.~\ref{fig:hist_b}) and column density (Fig.~\ref{fig:hist_Nh}) on the top rows.   These quantities may be subject to systematic uncertainties in the reconstruction of the pathlength but  should be less sensitive to uncertainties in including narrow or low-column density lines in the fits. We also show the probability distribution functions of the linewidth and column density (bottom rows), which are unaffected by uncertainties on the pathlength but their normalization can be affected by numerical resolution limits, difficulties in reliably fitting narrow or low-column density lines, and errors in continuum placement. Line fitting algorithms sometimes fail to identify and/or model lines with a low column density ($\log(N_{\mathrm{HI}})<12$). We have, hence, shaded the corresponding area in our plots and results should be considered with caution there. 

The distribution of the linewidths shows that the heated models lack narrow lines (below 25 km s$^{-1}$),  while all models have similar numbers of broader lines. This leads to a global shift of the PDF towards broader lines for heated models (lower row). The unheated model reproduces both the normalised and un-normalised data, while the heated models fail at reproducing them.
The distribution of column densities show that heated models underproduce the total number of lines by roughly 50$\%$ near the peak
of the distribution and that they somewhat overproduce systems with higher column densities. The unheated model reproduces both the normalised and un-normalised data, while the heated models fail at reproducing them.

\section{Discussion}\label{sec:discussion}

In P12 we showed that observations of the Lyman $\alpha$ forest such as the distribution of linewidths and column densities as well as the flux PDF and powerspectrum around redshift 3 were in good agreement with predictions accounting for blazar heating. In this paper, we find that i) a rescaled flux PDF from a single high-resolution spectrum from \citet{2017MNRAS.466.2690R} prefers models with inhomogeneous blazar heating at $z \sim 2.5-3$ and that  ii) the line-width column density distribution from \citet{Rudie_12_CGM} is not compatible with blazar heating at $z \sim 2.3$. At first glance these facts seem in contradiction with each other, but it is important to keep in mind that these two measurements probe not only different redshifts, but also different densities.
 The goal of this section is to explain how these statements can be reconciled  and how systematic uncertainties typically impact this type of analyses. We will discuss whether heating with a different redshift and/or density dependence (e.g.,  inhomogeneous HeII photoheating) could play a role and how to reconcile these results with evidence for blazar heating from other channels such as the gamma-ray sky.

\subsection{Systematic uncertainties and comparison with P12}

\begin{figure*}
\centering
\includegraphics[width = .95\textwidth ]{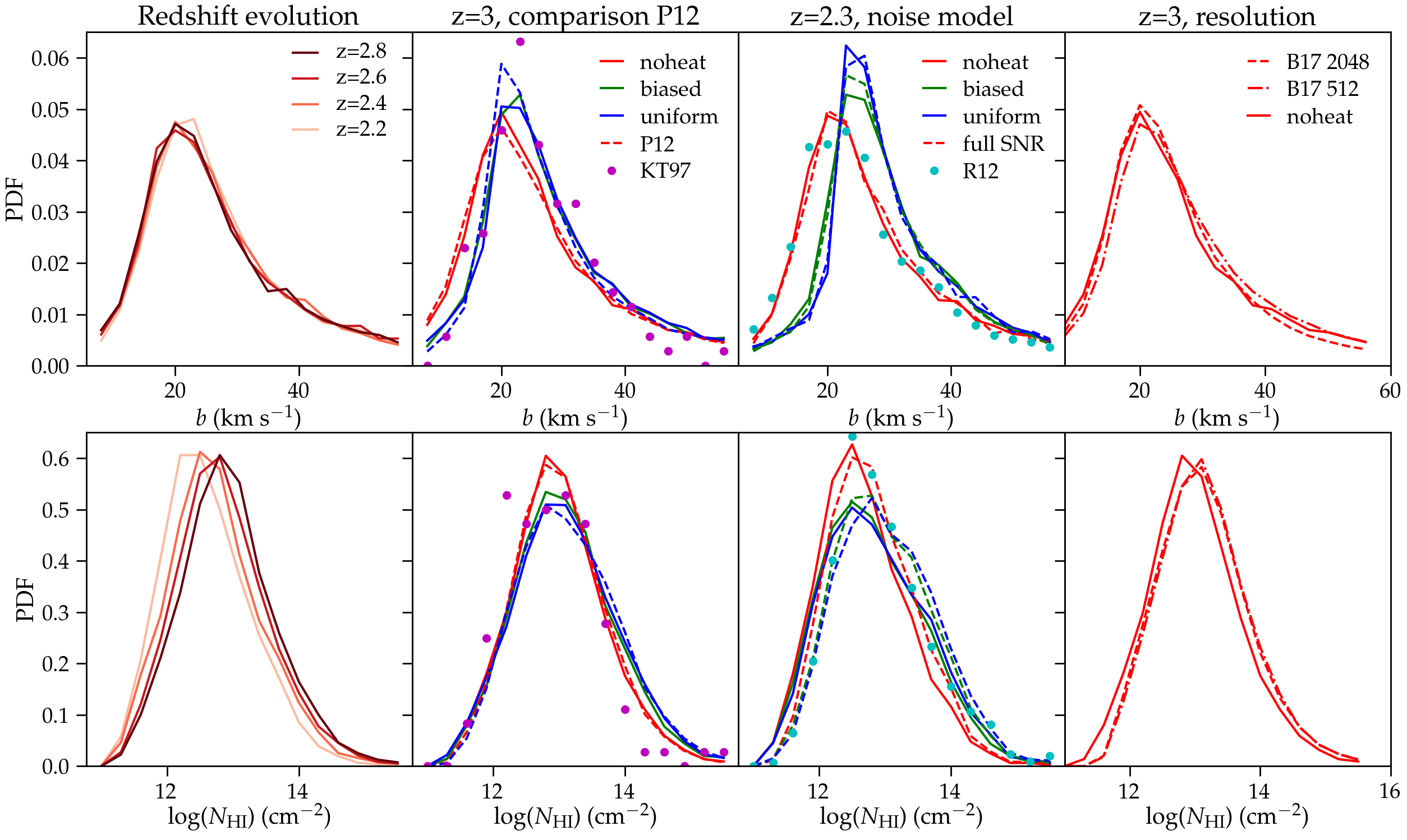}
\caption{Illustration of systematic effects impacting line width (top row) and column density (bottom row) probability distribution functions. First column: redshift evolution, from light to deeper red in the unheated model. Second column: high redshift simulations show consistency with the P12 results and the heated model matches the \citet{1997ApJ...484..672K} data.  Third column: low-redshift simulations and impact of noise model. The unheated model is preferred by the ~\citet{Rudie_2012_T_rho} data. Final column: Impact of increased resolution  illustrated by the Sherwood simulations. }
\label{fig:systematics}
\end{figure*}

Here, we detail redshift sampling effects, signal-to-noise models, and resolution effects, all of which are illustrated in the linewidth and column density PDFs shown in Fig.~\ref{fig:systematics}.

The left panel of Fig.~\ref{fig:systematics} illustrates how the linewidth and column density distributions evolve with redshift (darker red means increasing redshift). In the unheated model shown here, the whole column density distribution shifts by about half a  dex  between $z=2$ and $z=2.8$ (reflecting the significant evolution in the mean cosmic density and the opacity of the Lyman $\alpha$ forest), while the line width distribution is mostly unchanged (suggesting less evolution in absorber temperature\footnote{Note, however, that lines of fixed column density probe different overdensities at different redshifts.}). In the inhomogeneous heating model (not shown here) the linewidth distribution is globally shifted by 5 km s$^{-1}$ towards higher velocities at low redshift. These changes due to the redshift evolution are of comparable magnitude than those resulting from different heating models. Often observational results cover a range of redshifts, and the actual redshift distribution of the individual absorbers can vary significantly. Any comparison between observations and numerical simulations should therefore properly reconstruct the pathlength distribution, at the risk of otherwise introducing systematic shifts in the results. 

In P12, the $z=3$ simulated $b-N_{\mathrm{HI}}$ distribution was compared with the observations from \citet{1997ApJ...484..672K} between $2.75<z<3.05$. The results are recalled here in the second panel of Fig.~\ref{fig:systematics}, showing that the observed distribution (purple dots) and uniform heating model (dashed blue line) are in good agreement. We also show that the simulations presented here (solid lines) are in good agreement with the simulations from P12 (dashed lines), for all the  models. As such, the simulations presented here are fully compatible with the ones from P12. We show the less informative powerspectrum and flux PDF in Appendix A for completeness. The need for additional heating found in the $b-N_{\mathrm{HI}}$ distribution around $z\simeq$ 3 is in agreement with our analysis of the rescaled flux PDF of \citet{2017MNRAS.466.2690R}, which covered the redshift interval $[2.5-3.0]$.

In this publication we considered a lower redshift region for the $b-N_{\mathrm{HI}}$ distribution, focusing on the more recent data by  \citet{Rudie_2012_T_rho} between $2.0<z<2.8$. We modeled a fully reconstructed redshift distribution from the simulation and find that our blazar heated models do not agree with the data. If we restrict ourselves to  the redshift range in which most of the data lies ($2.2<z<2.4$) and compare to the $z=2.3$ snapshot, we find a qualitatively similar result as in our main analysis, as is shown in the third column of Fig.~\ref{fig:systematics}.  While  our results and the P12 results seem incompatible at first, this discrepancy can be explained by different redshifts in consideration. At high redshift ($z\simeq 3$), additional heating of the IGM is needed, while at lower redshift ($z\simeq 2.3$) additional heating is  strongly disfavoured by the data. This point will be discussed further in Sec.~\ref{sec:heating_evolution}.

The evolution of the SNR is another subtlety relating to redshift. Results are typically presented with a mean SNR but important variations can be present. As an example, the data in \citet{Rudie_12_CGM} has a typical SNR of 100 but high redshift lines can have an SNR as high as 250 and low redshift lines can go as low as 50. We have modeled the redshift-dependent noise-distribution to match the observational data. The dashed lines in the third column of Fig.~\ref{fig:systematics} show the distributions with the complete noise model in comparison with the average SNR $=100$ model at $z=2.3$. At this redshift, which is the mode of the redshift distribution, the difference is minimal and does not affect the comparison between simulations and observations presented here. However, we emphasize that this should be checked systematically, especially when one focuses on subsets of the data, which may have a systematically smaller/larger SNR than the bulk of the data.

Finally, the last row of Fig.~\ref{fig:systematics} shows the impact of numerical resolution on the line parameters, based on the Sherwood simulation data at $z=3$. Given the large size of the regions heated by blazars (see Fig.~\ref{fig:T_flucs}), our simulation domain covers 100$\,h^{-1}$Mpc on the side, which limits its effective mass resolution. ~\citet{Bolton_17_sherwood} present a detailed study of resolution effects, based on their high resolution data (2048$^3$ particles with a boxlength of 40 h$^{-1} $Mpc on the side, dashed line) and a lower resolution simulation (512$^{3}$ particles, dash-dotted line), which has an effective resolution similar to ours. The increased resolution results in narrower lines, shifted by a few km s$^{-1}$ and has almost no impact on the PDF of the column densities considered here.  A similar study at $z=2$ (not shown here) shows a smaller impact of resolution. Globally, resolution effects are smaller than the differences between our different heating models and are not expected to affect the conclusions of this work significantly. 

Several additional systematic uncertainties can affect this work, and we have checked that their impact is smaller than the differences between the heating models. Our 100 h$^{-1}$ Mpc on the size simulation domain provides a good representation of large scale structure in the Universe, up to group sized objects, which could hosts some of the blazars we are considering, and the impact of cosmic variance should be limited. The 15 lines-of-sight composing the \citet{Rudie_2012_T_rho} data also provide a good coverage of cosmic variance. On the other hand, the~\citet{2017MNRAS.466.2690R} study is only based on a single line of sight, and effects of cosmic variance may be important. To mitigate this we have bootstrapped our data and computed 100 mock spectra, allowing for a meaningful comparison. Additionally, as shown in Figs.~\ref{fig:hist_b}-\ref{fig:hist_Nh}, the uncertainty due to metal contamination in observational data is small. Also, we have tested that there are only minor differences in line properties between the line-fitting algorithm VPFIT which we have used here and AUTOVP \citep{Dave_1997_autovp} which was used in P12. Finally, errors in the continuum placement can affect the comparison between our models and the line properties from \citet{Rudie_2012_T_rho}. Given the high SNR of the observations, only absorbers with log($N_{HI}$)<12.5 would be affected by this. This could potentially slightly skew the PDF of the linewidths and/or column densities but would let the un-normalized distributions mostly unaffected (top rows of Figs.~\ref{fig:hist_b}-\ref{fig:hist_Nh}). Specifically, the strong preference for the unheated model in the distribution of the column density will be unchanged. In conclusion, we find no systematic uncertainties which  could qualitatively change our main conclusions.

\subsection{Redshift evolution and physical mechanism of IGM heating}
\label{sec:heating_evolution}

The seemingly contradictory results found in this study could be reconciled if significant heating of the low-density IGM occurs at $z\gtrsim3$, and if the IGM cools down towards lower redshifts. This redshift trend is, however, unexpected for heating by TeV blazars (see Fig.~\ref{fig:T0}).  

Photoheating associated with the reionisation of HeII is instead expected to deposit energy into the IGM during the epoch of HeII reionisation \citep[$z\gtrsim2.7$,][]{2016ApJ...825..144W}, while the IGM is expected to cool in the absence of blazar heating due to Hubble expansion and Compton cooling once helium is (almost) fully ionized. Compared to our simulations, homogeneous HeII heating is modelled more accurately in the Sherwood-Relics simulations, which use a non-equilibrium ionization/heating solver as well as a cosmic UV background that has been carefully matched to observational constraints on the reionisation history \citep{2019MNRAS.485...47P}. The corresponding thermal evolution of the IGM is illustrated in Fig.~\ref{fig:T0}. A temperature maximum is reached at $z\sim3$, while the temperature falls off at lower redshifts. Indeed the thermal evolution at mean density is similar to our blazar-heated models at $z\gtrsim3.5$ and close to our model without blazar heating at $z\lesssim2$. At the redshifts probed by the rescaled flux PDF of \citet{2017MNRAS.466.2690R}, the temperature in this model with increased HeII heating falls roughly half way between our models with and without blazar heating (although somewhat closer to the latter at the low densities, $\Delta \lesssim 0.5$, the rescaled flux PDF is most sensitive to). This is reflected in the rescaled flux PDF it predicts (gray curve in the left panel of Fig.~\ref{fig:PDF_A}). The deviation from the data is only about half of that in our simulation without blazar heating. Our inhomogeneously blazar-heated model is, however, in significantly better agreement with the data.

Nevertheless, these findings suggest that a model in which HeII heating occurs at similar redshifts as in the Sherwood-Relics simulations, but with an enhanced amplitude, might match all the data sets considered in this work. Additional heating would be primarily needed in the underdense gas probed by the rescaled flux PDF, while the temperature of denser gas probed by other techniques seems to be in better agreement with observational constraints \citep[see, e.g., ][]{Gaikwad_2020}.
A much larger amount of HeII heating is, however, not expected in most models, e.g., all the models considered in \citet{2019MNRAS.485...47P} and \citet{2020arXiv200205733U} predict temperatures at mean density that are at least a few thousand Kelvin lower at $z \sim 3$ then the blazar-heated models considered here.
In principle, radiative transfer effects during HeII reionization can specifically boost the temperature in the lowest density regions as they are often reionized last and have hence the least time to cool afterwards.
The most extreme radiative transfer model in \citet{2017ApJ...841...87L} (their simulation H5) comes very close in the temperature at mean density and has a fairly flat (but no inverted) temperature-density relation at lower densities. It is, hence, at this point not entirely clear whether the high IGM temperatures at low densities favoured by the rescaled flux PDF data can be realistically reached with an efficient HeII heating, or whether an additional heating channel such as provided by TeV blazars is needed.

\begin{figure*}
\centering
\includegraphics[width = .9\textwidth ]{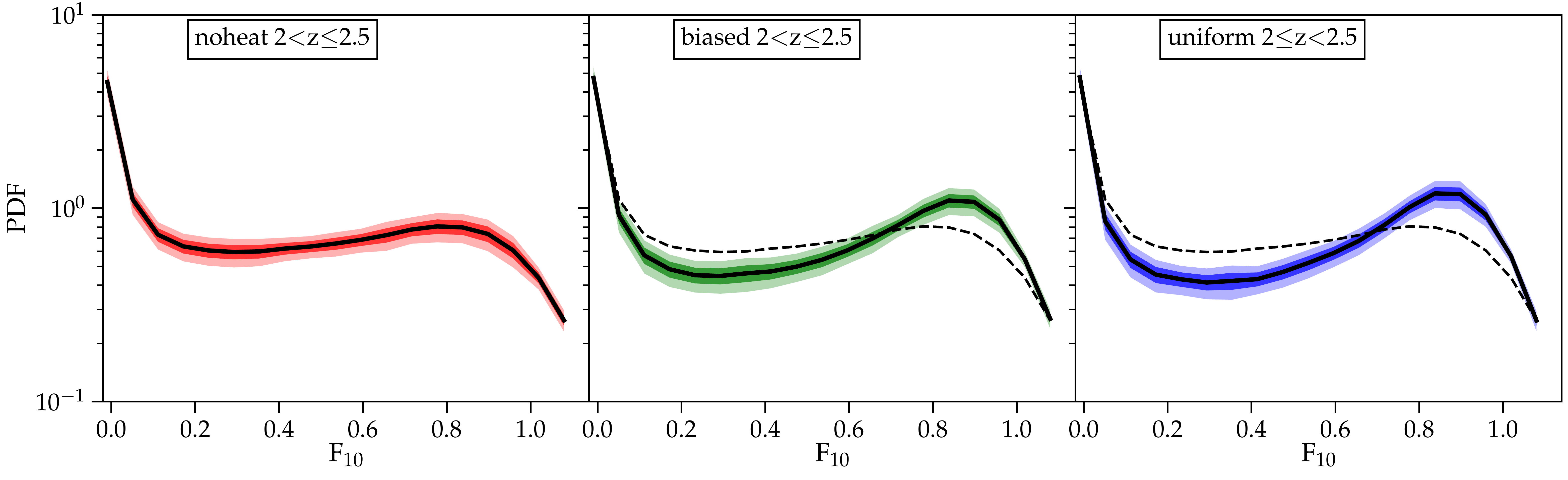}
\caption{Predicted probability distribution function of the rescaled flux in the unheated (left), inhomogeneously heated (middle) and uniformly heated model (right) between $2 <z\leq 2.5$. The thick line represents the mean over the 100 lines of sight, and the red and pink area the 1 and 2 $\sigma$ uncertainty around the mean. The corresponding numerical values are provided in Tab.~\ref{tab:predict_PDF} in the Appendix. The dashed line in the center and right plot show the mean in the unheated model for easier comparison.}
\label{fig:PDF_predict}
\end{figure*}

Fig.~\ref{fig:T0} illustrates that the difference in IGM temperature between models with and without blazar heating increases strongly at $z \lesssim 2.5$. Furthermore, blazar heating has the largest impact at the lowest densities (see Fig.~\ref{fig:rho_T}). Hence, we would ideally like to probe the IGM thermal state in underdense regions at lower redshifts. Unfortunately, the hydrogen Ly$\alpha$ forest becomes almost completely transparent there. Prominent Ly$\alpha$ absorbers correspond to increasingly dense systems at lower redshifts. Getting the maximum sensitivity to low density regions requires very high signal-to-noise data and analyzing regions that are almost transparent, which is essentially what was done in the analysis of \citet{2017MNRAS.466.2690R} for higher redshifts. Repeating such an analysis with similarly good data at $z\sim2$ to 2.5 should be a very promising way of constraining blazar heating with increased sensitivity. It would also be important to reduce the cosmic variance in the observed data set by using more lines of sight, compared to the single one that was studied in \citet{2017MNRAS.466.2690R}. In Fig.~\ref{fig:PDF_predict}, we show predictions of the rescaled flux PDF for this redshift range the data is provided in Table~\ref{tab:predict_PDF} in the Appendix for possible comparison with future observations or simulations. As expected the difference between the different heating models is larger at these lower redshifts. Data from future observational studies could be compared to this.

Another interesting approach would be to use HeII rather than hydrogen for absorption line studies of the thermal state of the IGM. The Ly$\alpha$ line of HeII also produces a forest of absorption lines, however, with a different redshift evolution and density dependence. Despite the lower abundance of helium compared to hydrogen, the HeII Ly$\alpha$ forest is more opaque due to the smaller number of photons that can ionize HeII, as well as due to the faster recombination of HeII. Unfortunately, the HeII Ly$\alpha$ forest needs to be observed from space and requires very bright background quasars without intervening Lyman limit systems that would absorb the relevant part of the spectrum. There is, hence, much less data available with typically lower signal-to-noise ratios. Nevertheless, it would be interesting to explore in a future work to what extent it can constrain blazar heating of the IGM.

\subsection{A unified picture of blazar heating: Ly$\alpha$ forest, gamma-ray sky and clustering}

Blazar heating results from heating of the IGM by plasma instabilities of electron-positron beams created by pair production from TeV gamma-rays. There are strong lines of evidence from gamma-ray observations that plasma instabilities drain the kinetic energy of electron-positron pairs that are produced by very high energy photon scattering off of $\sim 1$ eV photons from the extragalactic background light. 
In particular, an otherwise expected component of secondary inverse Compton-scattered gamma-rays is not observed, neither directly toward the BL-Lac sources nor when observing potential pair beams from the side. This limits intergalactic magnetic fields, which have been invoked to explain a deflection of the secondary component out of the beam direction, both from below and above, respectively. Limits of $B\gtrsim 10^{-14}-10^{-15}$ G \citep[see for instance][]{2010Sci...328...73N,2018ApJS..237...32A} and $B \lesssim 10^{-15}$ - $10^{-16}$ G \citep{2018ApJ...868...87B,2020ApJ...892..123T} have been derived, where the latter use a novel analysis that explores the anisotropic nature of the gamma-ray halos \citep{2016ApJ...832..109B,2017ApJ...850..157T}. These mutually exclusive constraints preclude magnetic deflection of pairs as a mechanism of suppression of the secondary inverse Compton emission and suggest that plasma instabilities may instead be at work.  

In addition, the principle plasma instability that is driven by these pair beams, the so-called ``oblique instability'' \citep{PhysRevE.70.046401,2012ApJ...752...22B,2016ApJ...833..118C} appears to be robust in the conditions of the intergalactic medium against both nonlinear effects \citep{2014ApJ...797..110C} and large scale density gradients \citep{2018ApJ...859...45S,2020JPlPh..86b5301S}. This is not to say that the principle instability that is driven by pair beams is the ``oblique instability'', but generic arguments for the suppression of plasma instabilities are erroneous \citep{2013ApJ...770...54M}.

The results of the observed {\it lack} of blazar heating in the Ly$\alpha$ forest at  $z\lesssim2.5$ are in tension with the lines of evidence from the gamma-ray sky.  One possibility is that the plasma unstable modes that are driven by these pair beams do not ultimately end up as thermal heat, but remain in small scale magnetic fields and plasma waves. The extremely-long-term evolution of these modes have never been studied and much of our intuition of the behavior of these plasma instabilities are interpolated from a much more prosaic region of parameter space.

Another possibility is that the redshift evolution for blazar heating is different from what we have assumed in this work and that the blazar luminosity function may drop off faster toward high redshift, $z\gtrsim1.5-2$. In our previous work, we rescaled the quasar luminosity function to produce the blazar luminosity function by noting that the blazar luminosity distribution appears to be a rescaling of the quasar luminosity distribution at $z=0.1$ \citep{2012ApJ...752...22B,2012ApJ...752...23C}. However, the evidence for the presence of IGM beam-plasma instabilities and the absence of the re-processed inverse Compton-scattered GeV gamma-ray photons accumulates mostly at low redshift ($z\lesssim1.5$), whereas the evidence from the Ly$\alpha$ is from high redshift ($z\gtrsim2.5$). 

In particular, the number of resolved TeV blazars probe very low redshifts ($z\lesssim 0.5$). We demonstrate in \citet{2014ApJ...790..137B} that a positively evolving distribution of hard gamma-ray blazars (which are similar to high-frequency synchrotron peaked BL Lacs) in combination with the presence of virulent plasma beam instabilities that preempt the inverse Compton cascade provides an excellent match to the observational data, in particular the hard gamma-ray blazar redshift distribution and $\log \mathcal{N}$-$\log S$ distribution (where $\mathcal{N}$ and $S$ are the number and flux of nearby hard gamma-ray-bright blazars in the {\em Fermi} gamma-ray band, respectively). Moreover, more than 90 per cent of the isotropic extragalactic gamma-ray background at energies $E_\gamma>10$~GeV \citep{2015ApJ...799...86A} is accounted for by hard gamma-ray blazars at redshifts $z\lesssim 2$ \citep[see figure 9 of][]{2014ApJ...790..137B} with the remainder likely contributed by starburst galaxies \citep{2012ApJ...755..164A}. The low-energy ($E_\gamma<10$~GeV) isotropic extragalactic gamma-ray background is likely dominated by unresolved soft gamma-ray blazars and flat-spectrum radio quasars (FSRQs), which are known to show a positive redshift evolution with increasing $z$ \citep{2014ApJ...780...73A}.

However, this fortuitous relationship between the blazar and quasar luminosity distribution \citep{2012ApJ...752...22B} may only exist at low redshift ($z\lesssim1.5$) and diverge at higher redshift.  One line of argument that this might be the case is that the analysis of the two-point correlation function of BL-Lacs and FSRQs from \textit{Fermi} suggest that these objects occupy dark matter halos of $\sim 2\times 10^{13}\, \rmn{M}_{\odot}$ at least up to $z\sim 1$ \citep{2014ApJ...797...96A}.  This is inline with the halos hosting radio loud SDSS quasars between $z=0.3-1.3$ \citep{2009ApJ...697.1656S,2012ApJ...746....1K}.  On the other hand, optical quasars are associated with dark matter haloes of $\sim 10^{12}\, \rmn{M}_{\odot}$ \citep{2009ApJ...697.1656S,2014ApJ...797...96A}.  Because more massive dark matter halos that host BL-Lacs and FSRQs form later in a hierarchically growing universe \citep{2002ApJ...568...52W} and in consequence are much rarer at higher redshift, this would suggest that TeV emission and, hence, blazar heating falls off rapidly at higher redshift.

\section{Conclusions}\label{sec:conclusion}
We have performed  an in-depth comparison between  mock data from large-scale cosmological simulations from different blazar heating models and observations of the Ly$\alpha$ forest between $2<z<3$. 
In addition to previous work by \citet{Puchwein_12_Lya}, we have considered a more physically motivated inhomogeneous blazar heating model aside from the uniform heating model. This work was also motivated by a recently published very high resolution rescaled flux PDF for $2.5<z<3$ \citep{2017MNRAS.466.2690R} and a set of high resolution quasar absorption lines around $\langle z \rangle = 2.3$ \citep{Rudie_13_CGM}. The latter enables us to compare our simulations  with lower redshift observations than was done in P12. We find that:

\begin{itemize}
    \item Inhomogeneous blazar heating is in very good agreement with observations of the low-density IGM at redshift $2.5<z<3$. We base this on a comparison between our simulations and a very high resolution spectrum where the global optical depth was rescaled to enhance the contribution from underdense regions \citep{2017MNRAS.466.2690R}, which are most sensitive to blazar heating. Additional rescaling of the global transmitted flux also allowed  mitigating continuum placement uncertainties, thereby making this measurement possible.
    \item The linewidth and column density distributions determined by \citet{Rudie_13_CGM}  at redshift $\langle z \rangle = 2.3$ are incompatible with our blazar heated models and agree well with our model without additional heating.
    \item At first glance, our results seem in tension with P12, which concluded in favor of the presence of blazar heating. This can be explained by the different redshifts that were being considered in both publications. Most of the data favoring blazar heating in P12 was at $z>2.5$, while we perform a more detailed comparison to lower redshift ($z<2.5$) data here. We show that both results are not incompatible and  that carefully matching the redshift sampling of predictions to the observational data is important.
    \item We discuss whether heating from HeII reionisation alone could explain the rescaled flux PDF data. While the redshift evolution required for the heating is consistent with expectations for HeII photoheating, the amount of heating is insufficient for most models of HeII reionisation. It may work, however, for the most extreme models present in the literature, in particular if radiative transfer effects can enhance heating of low density regions.
    \item Overall, the comparisons from this paper call for a modification of blazar heating of the IGM as described in \citet{2012ApJ...752...22B,2012ApJ...752...23C, 2012ApJ...752...24P}. However, observations of the gamma-ray sky and theoretical work on plasma instabilities does provide evidence of plasma instabilities at work. Whether the energy is eventually transferred through other channels than heating or whether the evolutionary history of blazar heating needs to be revised remains a question. Clustering studies of blazars provide a hint that the blazar luminosity redshift distribution decreases faster towards high redshift in comparison to the quasar luminosity distribution, thus providing circumstantial evidence that the latter explanation may at least explain part of the effect found in this study.
    
\end{itemize}

Our thorough comparison between mock data and observations has provided us with new insight on blazar heating and the thermal history of the IGM in a broader sense. A firmer answer on (a potentially delayed) blazar heating will rely on additional data, covering a wider range of redshifts. We particularly advocate for an analysis similar to \citet{2017MNRAS.466.2690R} focused on lower redshift ($z<2.5$). Studies of the HeII Lyman-$\alpha$ forest may be another interesting avenue to explore to get a better handle on the thermal state of the very low density IGM. In all cases, we find that discriminating comparisons between simulations and observations are only possible when both datasets are well described and documented, thus reducing systematic uncertainties (e.g. redshift sampling, signal-to-noise model...).

\section*{Acknowledgements}
The authors warmly thank James Bolton for his insightful comments on this work, as well as for making line lists from the Sherwood simulations available to us. A.L. thank Alberto Rorai for  many clarifications he provided on his work.
CP and MS acknowledge support by the European Research Council under ERC-CoG grant CRAGSMAN-646955.
PC acknowledges support by the NASA ATP program through NASA grant NNH17ZDA001N-ATP.
This work was supported in part by Perimeter Institute for Theoretical Physics.  Research at Perimeter Institute is supported by the Government of Canada through the Department of Innovation, Science and Economic Development Canada and by the Province of Ontario through the Ministry of Economic Development, Job Creation and Trade.
PT receives support from the Natural Science and Engineering Research Council through the Alexander Graham Bell CGS-D scholarship.
A.E.B. thanks the Delaney Family for their generous financial support via the Delaney Family John A. Wheeler Chair at Perimeter Institute.
A.E.B. receives additional financial support from the Natural Sciences and Engineering Research Council of Canada through a Discovery Grant.
The Sherwood and Sherwood-Relics simulations were performed using the Curie and Irene supercomputers at the Tres Grand Centre de Calcul (TGCC), and the DiRAC Data Analytic system at the University of Cambridge. The latter is operated by the University of Cambridge High Performance Computing Service on behalf of the STFC DiRAC HPC Facility (www.dirac.ac.uk). This equipment was funded by BIS National E-infrastructure capital grant (ST/K001590/1), STFC capital grants ST/H008861/1 and ST/H00887X/1, and STFC DiRAC Operations grant ST/K00333X/1. DiRAC is part of the National E-Infrastructure.

\section*{Data Availability}
 The data underlying this article will be shared on reasonable request to the corresponding author.




\bibliographystyle{mnras}
\bibliography{biblio_total} 

\begin{thebibliography}{}
\makeatletter
\relax
\def\mn@urlcharsother{\let\do\@makeother \do\$\do\&\do\#\do\^\do\_\do\%\do\~}
\def\mn@doi{\begingroup\mn@urlcharsother \@ifnextchar [ {\mn@doi@}
  {\mn@doi@[]}}
\def\mn@doi@[#1]#2{\def\@tempa{#1}\ifx\@tempa\@empty \href
  {http://dx.doi.org/#2} {doi:#2}\else \href {http://dx.doi.org/#2} {#1}\fi
  \endgroup}
\def\mn@eprint#1#2{\mn@eprint@#1:#2::\@nil}
\def\mn@eprint@arXiv#1{\href {http://arxiv.org/abs/#1} {{\tt arXiv:#1}}}
\def\mn@eprint@dblp#1{\href {http://dblp.uni-trier.de/rec/bibtex/#1.xml}
  {dblp:#1}}
\def\mn@eprint@#1:#2:#3:#4\@nil{\def\@tempa {#1}\def\@tempb {#2}\def\@tempc
  {#3}\ifx \@tempc \@empty \let \@tempc \@tempb \let \@tempb \@tempa \fi \ifx
  \@tempb \@empty \def\@tempb {arXiv}\fi \@ifundefined
  {mn@eprint@\@tempb}{\@tempb:\@tempc}{\expandafter \expandafter \csname
  mn@eprint@\@tempb\endcsname \expandafter{\@tempc}}}

\bibitem[\protect\citeauthoryear{{Ackermann} et~al.,}{{Ackermann}
  et~al.}{2012}]{2012ApJ...755..164A}
{Ackermann} M.,  et~al., 2012, \mn@doi [\apj] {10.1088/0004-637X/755/2/164},
  \href {https://ui.adsabs.harvard.edu/abs/2012ApJ...755..164A} {755, 164}

\bibitem[\protect\citeauthoryear{{Ackermann} et~al.,}{{Ackermann}
  et~al.}{2015}]{2015ApJ...799...86A}
{Ackermann} M.,  et~al., 2015, \mn@doi [\apj] {10.1088/0004-637X/799/1/86},
  \href {https://ui.adsabs.harvard.edu/abs/2015ApJ...799...86A} {799, 86}

\bibitem[\protect\citeauthoryear{{Ackermann} et~al.,}{{Ackermann}
  et~al.}{2018}]{2018ApJS..237...32A}
{Ackermann} M.,  et~al., 2018, \mn@doi [\apjs] {10.3847/1538-4365/aacdf7},
  \href {https://ui.adsabs.harvard.edu/abs/2018ApJS..237...32A} {237, 32}

\bibitem[\protect\citeauthoryear{{Ajello} et~al.,}{{Ajello}
  et~al.}{2014}]{2014ApJ...780...73A}
{Ajello} M.,  et~al., 2014, \mn@doi [\apj] {10.1088/0004-637X/780/1/73}, \href
  {https://ui.adsabs.harvard.edu/abs/2014ApJ...780...73A} {780, 73}

\bibitem[\protect\citeauthoryear{{Allevato}, {Finoguenov}  \&
  {Cappelluti}}{{Allevato} et~al.}{2014}]{2014ApJ...797...96A}
{Allevato} V.,  {Finoguenov} A.,   {Cappelluti} N.,  2014, \mn@doi [\apj]
  {10.1088/0004-637X/797/2/96}, \href
  {https://ui.adsabs.harvard.edu/abs/2014ApJ...797...96A} {797, 96}

\bibitem[\protect\citeauthoryear{{Becker} \& {Bolton}}{{Becker} \&
  {Bolton}}{2013}]{2013MNRAS.436.1023B}
{Becker} G.~D.,  {Bolton} J.~S.,  2013, \mn@doi [\mnras]
  {10.1093/mnras/stt1610}, \href
  {http://adsabs.harvard.edu/abs/2013MNRAS.436.1023B} {436, 1023}

\bibitem[\protect\citeauthoryear{{Becker}, {Bolton}, {Haehnelt}  \&
  {Sargent}}{{Becker} et~al.}{2011}]{2011MNRAS.410.1096B}
{Becker} G.~D.,  {Bolton} J.~S.,  {Haehnelt} M.~G.,   {Sargent} W.~L.~W.,
  2011, \mn@doi [\mnras] {10.1111/j.1365-2966.2010.17507.x}, \href
  {http://adsabs.harvard.edu/abs/2011MNRAS.410.1096B} {410, 1096}

\bibitem[\protect\citeauthoryear{{Becker}, {Hewett}, {Worseck}  \&
  {Prochaska}}{{Becker} et~al.}{2013}]{Becker_13_tau}
{Becker} G.~D.,  {Hewett} P.~C.,  {Worseck} G.,   {Prochaska} J.~X.,  2013,
  \mn@doi [\mnras] {10.1093/mnras/stt031}, \href
  {https://ui.adsabs.harvard.edu/abs/2013MNRAS.430.2067B} {430, 2067}

\bibitem[\protect\citeauthoryear{{Boera}, {Murphy}, {Becker}  \&
  {Bolton}}{{Boera} et~al.}{2014}]{2014MNRAS.441.1916B}
{Boera} E.,  {Murphy} M.~T.,  {Becker} G.~D.,   {Bolton} J.~S.,  2014, \mn@doi
  [\mnras] {10.1093/mnras/stu660}, \href
  {http://adsabs.harvard.edu/abs/2014MNRAS.441.1916B} {441, 1916}

\bibitem[\protect\citeauthoryear{{Bolton}, {Viel}, {Kim}, {Haehnelt}  \&
  {Carswell}}{{Bolton} et~al.}{2008}]{2008MNRAS.386.1131B}
{Bolton} J.~S.,  {Viel} M.,  {Kim} T.-S.,  {Haehnelt} M.~G.,   {Carswell}
  R.~F.,  2008, \mn@doi [\mnras] {10.1111/j.1365-2966.2008.13114.x}, \href
  {http://adsabs.harvard.edu/abs/2008MNRAS.386.1131B} {386, 1131}

\bibitem[\protect\citeauthoryear{{Bolton}, {Becker}, {Haehnelt}  \&
  {Viel}}{{Bolton} et~al.}{2014}]{2014MNRAS.438.2499B}
{Bolton} J.~S.,  {Becker} G.~D.,  {Haehnelt} M.~G.,   {Viel} M.,  2014, \mn@doi
  [\mnras] {10.1093/mnras/stt2374}, \href
  {http://adsabs.harvard.edu/abs/2014MNRAS.438.2499B} {438, 2499}

\bibitem[\protect\citeauthoryear{{Bolton}, {Puchwein}, {Sijacki}, {Haehnelt},
  {Kim}, {Meiksin}, {Regan}  \& {Viel}}{{Bolton}
  et~al.}{2017}]{Bolton_17_sherwood}
{Bolton} J.~S.,  {Puchwein} E.,  {Sijacki} D.,  {Haehnelt} M.~G.,  {Kim} T.-S.,
   {Meiksin} A.,  {Regan} J.~A.,   {Viel} M.,  2017, \mn@doi [\mnras]
  {10.1093/mnras/stw2397}, \href
  {http://cdsads.u-strasbg.fr/abs/2017MNRAS.464..897B} {464, 897}

\bibitem[\protect\citeauthoryear{Bret, Firpo  \& Deutsch}{Bret
  et~al.}{2004}]{PhysRevE.70.046401}
Bret A.,  Firpo M.-C.,   Deutsch C.,  2004, \mn@doi [Phys. Rev. E]
  {10.1103/PhysRevE.70.046401}, 70, 046401

\bibitem[\protect\citeauthoryear{{Broderick}, {Chang}  \&
  {Pfrommer}}{{Broderick} et~al.}{2012}]{2012ApJ...752...22B}
{Broderick} A.~E.,  {Chang} P.,   {Pfrommer} C.,  2012, \mn@doi [\apj]
  {10.1088/0004-637X/752/1/22}, \href
  {http://adsabs.harvard.edu/abs/2012ApJ...752...22B} {752, 22}

\bibitem[\protect\citeauthoryear{{Broderick}, {Pfrommer}, {Puchwein}  \&
  {Chang}}{{Broderick} et~al.}{2014}]{2014ApJ...790..137B}
{Broderick} A.~E.,  {Pfrommer} C.,  {Puchwein} E.,   {Chang} P.,  2014, \mn@doi
  [\apj] {10.1088/0004-637X/790/2/137}, \href
  {https://ui.adsabs.harvard.edu/abs/2014ApJ...790..137B} {790, 137}

\bibitem[\protect\citeauthoryear{{Broderick}, {Tiede}, {Shalaby}, {Pfrommer},
  {Puchwein}, {Chang}  \& {Lamberts}}{{Broderick}
  et~al.}{2016}]{2016ApJ...832..109B}
{Broderick} A.~E.,  {Tiede} P.,  {Shalaby} M.,  {Pfrommer} C.,  {Puchwein} E.,
  {Chang} P.,   {Lamberts} A.,  2016, \mn@doi [\apj]
  {10.3847/0004-637X/832/2/109}, \href
  {https://ui.adsabs.harvard.edu/abs/2016ApJ...832..109B} {832, 109}

\bibitem[\protect\citeauthoryear{{Broderick}, {Tiede}, {Chang}, {Lamberts},
  {Pfrommer}, {Puchwein}, {Shalaby}  \& {Werhahn}}{{Broderick}
  et~al.}{2018}]{2018ApJ...868...87B}
{Broderick} A.~E.,  {Tiede} P.,  {Chang} P.,  {Lamberts} A.,  {Pfrommer} C.,
  {Puchwein} E.,  {Shalaby} M.,   {Werhahn} M.,  2018, \mn@doi [\apj]
  {10.3847/1538-4357/aae5f2}, \href
  {https://ui.adsabs.harvard.edu/abs/2018ApJ...868...87B} {868, 87}

\bibitem[\protect\citeauthoryear{{Calura}, {Tescari}, {D'Odorico}, {Viel},
  {Cristiani}, {Kim}  \& {Bolton}}{{Calura} et~al.}{2012}]{2012MNRAS.422.3019C}
{Calura} F.,  {Tescari} E.,  {D'Odorico} V.,  {Viel} M.,  {Cristiani} S.,
  {Kim} T.-S.,   {Bolton} J.~S.,  2012, \mn@doi [\mnras]
  {10.1111/j.1365-2966.2012.20811.x}, \href
  {http://adsabs.harvard.edu/abs/2012MNRAS.422.3019C} {422, 3019}

\bibitem[\protect\citeauthoryear{{Carswell} \& {Webb}}{{Carswell} \&
  {Webb}}{2014}]{Carswell_VPFIT}
{Carswell} R.~F.,  {Webb} J.~K.,  2014, Astrophysics Source Code Library, \href
  {http://adsabs.harvard.edu/abs/2014ascl.soft08015C} {}

\bibitem[\protect\citeauthoryear{{Chang}, {Broderick}  \& {Pfrommer}}{{Chang}
  et~al.}{2012}]{2012ApJ...752...23C}
{Chang} P.,  {Broderick} A.~E.,   {Pfrommer} C.,  2012, \mn@doi [\apj]
  {10.1088/0004-637X/752/1/23}, \href
  {http://adsabs.harvard.edu/abs/2012ApJ...752...23C} {752, 23}

\bibitem[\protect\citeauthoryear{{Chang}, {Broderick}, {Pfrommer}, {Puchwein},
  {Lamberts}  \& {Shalaby}}{{Chang} et~al.}{2014}]{2014ApJ...797..110C}
{Chang} P.,  {Broderick} A.~E.,  {Pfrommer} C.,  {Puchwein} E.,  {Lamberts} A.,
    {Shalaby} M.,  2014, \mn@doi [\apj] {10.1088/0004-637X/797/2/110}, \href
  {http://adsabs.harvard.edu/abs/2014ApJ...797..110C} {797, 110}

\bibitem[\protect\citeauthoryear{{Chang}, {Broderick}, {Pfrommer}, {Puchwein},
  {Lamberts}, {Shalaby}  \& {Vasil}}{{Chang}
  et~al.}{2016}]{2016ApJ...833..118C}
{Chang} P.,  {Broderick} A.~E.,  {Pfrommer} C.,  {Puchwein} E.,  {Lamberts} A.,
   {Shalaby} M.,   {Vasil} G.,  2016, \mn@doi [\apj]
  {10.3847/1538-4357/833/1/118}, \href
  {http://adsabs.harvard.edu/abs/2016ApJ...833..118C} {833, 118}

\bibitem[\protect\citeauthoryear{{Compostella}, {Cantalupo}  \&
  {Porciani}}{{Compostella} et~al.}{2013}]{2013MNRAS.435.3169C}
{Compostella} M.,  {Cantalupo} S.,   {Porciani} C.,  2013, \mn@doi [\mnras]
  {10.1093/mnras/stt1510}, \href
  {http://adsabs.harvard.edu/abs/2013MNRAS.435.3169C} {435, 3169}

\bibitem[\protect\citeauthoryear{{Dav{\'e}}, {Hernquist}, {Weinberg}  \&
  {Katz}}{{Dav{\'e}} et~al.}{1997}]{Dave_1997_autovp}
{Dav{\'e}} R.,  {Hernquist} L.,  {Weinberg} D.~H.,   {Katz} N.,  1997, \mn@doi
  [\apj] {10.1086/303712}, \href
  {https://ui.adsabs.harvard.edu/abs/1997ApJ...477...21D} {477, 21}

\bibitem[\protect\citeauthoryear{{Fan}, {Carilli}  \& {Keating}}{{Fan}
  et~al.}{2006}]{2006ARA&A..44..415F}
{Fan} X.,  {Carilli} C.~L.,   {Keating} B.,  2006, \mn@doi [\araa]
  {10.1146/annurev.astro.44.051905.092514}, \href
  {http://adsabs.harvard.edu/abs/2006ARA%26A..44..415F} {44, 415}

\bibitem[\protect\citeauthoryear{{Faucher-Gigu{\`e}re}, {Lidz}, {Zaldarriaga}
  \& {Hernquist}}{{Faucher-Gigu{\`e}re} et~al.}{2009}]{2009ApJ...703.1416F}
{Faucher-Gigu{\`e}re} C.-A.,  {Lidz} A.,  {Zaldarriaga} M.,   {Hernquist} L.,
  2009, \mn@doi [\apj] {10.1088/0004-637X/703/2/1416}, \href
  {http://adsabs.harvard.edu/abs/2009ApJ...703.1416F} {703, 1416}

\bibitem[\protect\citeauthoryear{{Gaikwad}, {Srianand}, {Haehnelt}  \&
  {Choudhury}}{{Gaikwad} et~al.}{2020a}]{Gaikwad_2020}
{Gaikwad} P.,  {Srianand} R.,  {Haehnelt} M.~G.,   {Choudhury} T.~R.,  2020a,
  arXiv e-prints, \href {https://ui.adsabs.harvard.edu/abs/2020arXiv200900016G}
  {p. arXiv:2009.00016}

\bibitem[\protect\citeauthoryear{{Gaikwad} et~al.,}{{Gaikwad}
  et~al.}{2020b}]{2020MNRAS.494.5091G}
{Gaikwad} P.,  et~al., 2020b, \mn@doi [\mnras] {10.1093/mnras/staa907}, \href
  {https://ui.adsabs.harvard.edu/abs/2020MNRAS.494.5091G} {494, 5091}

\bibitem[\protect\citeauthoryear{{Haardt} \& {Madau}}{{Haardt} \&
  {Madau}}{2012}]{2012ApJ...746..125H}
{Haardt} F.,  {Madau} P.,  2012, \mn@doi [\apj] {10.1088/0004-637X/746/2/125},
  \href {http://adsabs.harvard.edu/abs/2012ApJ...746..125H} {746, 125}

\bibitem[\protect\citeauthoryear{{Hiss}, {Walther}, {O{\~n}orbe}  \&
  {Hennawi}}{{Hiss} et~al.}{2019}]{Hiss_19_bNhfull}
{Hiss} H.,  {Walther} M.,  {O{\~n}orbe} J.,   {Hennawi} J.~F.,  2019, arXiv
  e-prints, \href {http://cdsads.u-strasbg.fr/abs/2019arXiv190311940H} {}

\bibitem[\protect\citeauthoryear{{Hopkins}, {Richards}  \&
  {Hernquist}}{{Hopkins} et~al.}{2007}]{2007ApJ...654..731H}
{Hopkins} P.~F.,  {Richards} G.~T.,   {Hernquist} L.,  2007, \mn@doi [\apj]
  {10.1086/509629}, \href {http://adsabs.harvard.edu/abs/2007ApJ...654..731H}
  {654, 731}

\bibitem[\protect\citeauthoryear{{Hui} \& {Gnedin}}{{Hui} \&
  {Gnedin}}{1997}]{1997MNRAS.292...27H}
{Hui} L.,  {Gnedin} N.~Y.,  1997, \mnras, \href
  {http://adsabs.harvard.edu/abs/1997MNRAS.292...27H} {292, 27}

\bibitem[\protect\citeauthoryear{{Hummels}, {Smith}  \& {Silvia}}{{Hummels}
  et~al.}{2017}]{Hummels_17_Trident}
{Hummels} C.~B.,  {Smith} B.~D.,   {Silvia} D.~W.,  2017, \mn@doi [\apj]
  {10.3847/1538-4357/aa7e2d}, \href
  {http://cdsads.u-strasbg.fr/abs/2017ApJ...847...59H} {847, 59}

\bibitem[\protect\citeauthoryear{{Kim}, {Bolton}, {Viel}, {Haehnelt}  \&
  {Carswell}}{{Kim} et~al.}{2007}]{2007MNRAS.382.1657K}
{Kim} T.-S.,  {Bolton} J.~S.,  {Viel} M.,  {Haehnelt} M.~G.,   {Carswell}
  R.~F.,  2007, \mn@doi [\mnras] {10.1111/j.1365-2966.2007.12406.x}, \href
  {http://adsabs.harvard.edu/abs/2007MNRAS.382.1657K} {382, 1657}

\bibitem[\protect\citeauthoryear{{Kirkman} \& {Tytler}}{{Kirkman} \&
  {Tytler}}{1997}]{1997ApJ...484..672K}
{Kirkman} D.,  {Tytler} D.,  1997, \mn@doi [\apj] {10.1086/304371}, \href
  {https://ui.adsabs.harvard.edu/abs/1997ApJ...484..672K} {484, 672}

\bibitem[\protect\citeauthoryear{{Krumpe}, {Miyaji}, {Coil}  \&
  {Aceves}}{{Krumpe} et~al.}{2012}]{2012ApJ...746....1K}
{Krumpe} M.,  {Miyaji} T.,  {Coil} A.~L.,   {Aceves} H.,  2012, \mn@doi [\apj]
  {10.1088/0004-637X/746/1/1}, \href
  {https://ui.adsabs.harvard.edu/abs/2012ApJ...746....1K} {746, 1}

\bibitem[\protect\citeauthoryear{{Kulkarni}, {Keating}, {Haehnelt}, {Bosman},
  {Puchwein}, {Chardin}  \& {Aubert}}{{Kulkarni}
  et~al.}{2019}]{2019MNRAS.485L..24K}
{Kulkarni} G.,  {Keating} L.~C.,  {Haehnelt} M.~G.,  {Bosman} S. E.~I.,
  {Puchwein} E.,  {Chardin} J.,   {Aubert} D.,  2019, \mn@doi [\mnras]
  {10.1093/mnrasl/slz025}, \href
  {https://ui.adsabs.harvard.edu/abs/2019MNRAS.485L..24K} {485, L24}

\bibitem[\protect\citeauthoryear{{La Plante}, {Trac}, {Croft}  \& {Cen}}{{La
  Plante} et~al.}{2017}]{2017ApJ...841...87L}
{La Plante} P.,  {Trac} H.,  {Croft} R.,   {Cen} R.,  2017, \mn@doi [\apj]
  {10.3847/1538-4357/aa7136}, \href
  {https://ui.adsabs.harvard.edu/abs/2017ApJ...841...87L} {841, 87}

\bibitem[\protect\citeauthoryear{{Lamberts}, {Chang}, {Pfrommer}, {Puchwein},
  {Broderick}  \& {Shalaby}}{{Lamberts} et~al.}{2015}]{2015ApJ...811...19L}
{Lamberts} A.,  {Chang} P.,  {Pfrommer} C.,  {Puchwein} E.,  {Broderick} A.~E.,
    {Shalaby} M.,  2015, \mn@doi [\apj] {10.1088/0004-637X/811/1/19}, \href
  {http://adsabs.harvard.edu/abs/2015ApJ...811...19L} {811, 19}

\bibitem[\protect\citeauthoryear{{Lee}}{{Lee}}{2012}]{2012ApJ...753..136L}
{Lee} K.-G.,  2012, \mn@doi [\apj] {10.1088/0004-637X/753/2/136}, \href
  {http://adsabs.harvard.edu/abs/2012ApJ...753..136L} {753, 136}

\bibitem[\protect\citeauthoryear{{Lynds}}{{Lynds}}{1971}]{1971ApJ...164L..73L}
{Lynds} R.,  1971, \mn@doi [\apjl] {10.1086/180695}, \href
  {http://adsabs.harvard.edu/abs/1971ApJ...164L..73L} {164, L73}

\bibitem[\protect\citeauthoryear{{McQuinn} \& {Upton Sanderbeck}}{{McQuinn} \&
  {Upton Sanderbeck}}{2016}]{2016MNRAS.456...47M}
{McQuinn} M.,  {Upton Sanderbeck} P.~R.,  2016, \mn@doi [\mnras]
  {10.1093/mnras/stv2675}, \href
  {http://adsabs.harvard.edu/abs/2016MNRAS.456...47M} {456, 47}

\bibitem[\protect\citeauthoryear{{McQuinn}, {Lidz}, {Zaldarriaga}, {Hernquist},
  {Hopkins}, {Dutta}  \& {Faucher-Gigu{\`e}re}}{{McQuinn}
  et~al.}{2009}]{2009ApJ...694..842M}
{McQuinn} M.,  {Lidz} A.,  {Zaldarriaga} M.,  {Hernquist} L.,  {Hopkins} P.~F.,
   {Dutta} S.,   {Faucher-Gigu{\`e}re} C.-A.,  2009, \mn@doi [\apj]
  {10.1088/0004-637X/694/2/842}, \href
  {http://adsabs.harvard.edu/abs/2009ApJ...694..842M} {694, 842}

\bibitem[\protect\citeauthoryear{{Meiksin} \& {Tittley}}{{Meiksin} \&
  {Tittley}}{2012}]{2012MNRAS.423....7M}
{Meiksin} A.,  {Tittley} E.~R.,  2012, \mn@doi [\mnras]
  {10.1111/j.1365-2966.2011.20380.x}, \href
  {http://adsabs.harvard.edu/abs/2012MNRAS.423....7M} {423, 7}

\bibitem[\protect\citeauthoryear{{Miniati} \& {Elyiv}}{{Miniati} \&
  {Elyiv}}{2013}]{2013ApJ...770...54M}
{Miniati} F.,  {Elyiv} A.,  2013, \mn@doi [\apj] {10.1088/0004-637X/770/1/54},
  \href {http://adsabs.harvard.edu/abs/2013ApJ...770...54M} {770, 54}

\bibitem[\protect\citeauthoryear{{Neronov} \& {Vovk}}{{Neronov} \&
  {Vovk}}{2010}]{2010Sci...328...73N}
{Neronov} A.,  {Vovk} I.,  2010, \mn@doi [Science] {10.1126/science.1184192},
  \href {https://ui.adsabs.harvard.edu/abs/2010Sci...328...73N} {328, 73}

\bibitem[\protect\citeauthoryear{{Palanque-Delabrouille}
  et~al.,}{{Palanque-Delabrouille} et~al.}{2013}]{2013A&A...559A..85P}
{Palanque-Delabrouille} N.,  et~al., 2013, \mn@doi [\aap]
  {10.1051/0004-6361/201322130}, \href
  {http://adsabs.harvard.edu/abs/2013A%26A...559A..85P} {559, A85}

\bibitem[\protect\citeauthoryear{{Palanque-Delabrouille}
  et~al.,}{{Palanque-Delabrouille} et~al.}{2015}]{2015JCAP...11..011P}
{Palanque-Delabrouille} N.,  et~al., 2015, \mn@doi [JCAP]
  {10.1088/1475-7516/2015/11/011}, \href
  {http://adsabs.harvard.edu/abs/2015JCAP...11..011P} {11, 011}

\bibitem[\protect\citeauthoryear{{Pfrommer}, {Chang}  \&
  {Broderick}}{{Pfrommer} et~al.}{2012}]{2012ApJ...752...24P}
{Pfrommer} C.,  {Chang} P.,   {Broderick} A.~E.,  2012, \mn@doi [\apj]
  {10.1088/0004-637X/752/1/24}, \href
  {https://ui.adsabs.harvard.edu/abs/2012ApJ...752...24P} {752, 24}

\bibitem[\protect\citeauthoryear{{Planck Collaboration} et~al.,}{{Planck
  Collaboration} et~al.}{2014}]{2014A&A...571A..16P}
{Planck Collaboration} et~al., 2014, \mn@doi [\aap]
  {10.1051/0004-6361/201321591}, \href
  {http://adsabs.harvard.edu/abs/2014A%26A...571A..16P} {571, A16}

\bibitem[\protect\citeauthoryear{{Puchwein}, {Pfrommer}, {Springel},
  {Broderick}  \& {Chang}}{{Puchwein} et~al.}{2012}]{Puchwein_12_Lya}
{Puchwein} E.,  {Pfrommer} C.,  {Springel} V.,  {Broderick} A.~E.,   {Chang}
  P.,  2012, \mn@doi [\mnras] {10.1111/j.1365-2966.2012.20738.x}, \href
  {http://cdsads.u-strasbg.fr/abs/2012MNRAS.423..149P} {423, 149}

\bibitem[\protect\citeauthoryear{{Puchwein}, {Bolton}, {Haehnelt}, {Madau},
  {Becker}  \& {Haardt}}{{Puchwein} et~al.}{2015}]{2015MNRAS.450.4081P}
{Puchwein} E.,  {Bolton} J.~S.,  {Haehnelt} M.~G.,  {Madau} P.,  {Becker}
  G.~D.,   {Haardt} F.,  2015, \mn@doi [\mnras] {10.1093/mnras/stv773}, \href
  {http://adsabs.harvard.edu/abs/2015MNRAS.450.4081P} {450, 4081}

\bibitem[\protect\citeauthoryear{{Puchwein}, {Haardt}, {Haehnelt}  \&
  {Madau}}{{Puchwein} et~al.}{2019}]{2019MNRAS.485...47P}
{Puchwein} E.,  {Haardt} F.,  {Haehnelt} M.~G.,   {Madau} P.,  2019, \mn@doi
  [\mnras] {10.1093/mnras/stz222}, \href
  {https://ui.adsabs.harvard.edu/abs/2019MNRAS.485...47P} {485, 47}

\bibitem[\protect\citeauthoryear{{Rafighi}, {Vafin}, {Pohl}  \&
  {Niemiec}}{{Rafighi} et~al.}{2017}]{2017A&A...607A.112R}
{Rafighi} I.,  {Vafin} S.,  {Pohl} M.,   {Niemiec} J.,  2017, \mn@doi [\aap]
  {10.1051/0004-6361/201731127}, \href
  {https://ui.adsabs.harvard.edu/abs/2017A&A...607A.112R} {607, A112}

\bibitem[\protect\citeauthoryear{{Rauch} et~al.,}{{Rauch}
  et~al.}{1997}]{1997ApJ...489....7R}
{Rauch} M.,  et~al., 1997, \mn@doi [\apj] {10.1086/304765}, \href
  {http://adsabs.harvard.edu/abs/1997ApJ...489....7R} {489, 7}

\bibitem[\protect\citeauthoryear{{Rollinde}, {Theuns}, {Schaye}, {P{\^a}ris}
  \& {Petitjean}}{{Rollinde} et~al.}{2013}]{2013MNRAS.428..540R}
{Rollinde} E.,  {Theuns} T.,  {Schaye} J.,  {P{\^a}ris} I.,   {Petitjean} P.,
  2013, \mn@doi [\mnras] {10.1093/mnras/sts057}, \href
  {http://adsabs.harvard.edu/abs/2013MNRAS.428..540R} {428, 540}

\bibitem[\protect\citeauthoryear{{Rorai} et~al.,}{{Rorai}
  et~al.}{2017}]{2017MNRAS.466.2690R}
{Rorai} A.,  et~al., 2017, \mn@doi [\mnras] {10.1093/mnras/stw2917}, \href
  {http://adsabs.harvard.edu/abs/2017MNRAS.466.2690R} {466, 2690}

\bibitem[\protect\citeauthoryear{{Rorai}, {Carswell}, {Haehnelt}, {Becker},
  {Bolton}  \& {Murphy}}{{Rorai} et~al.}{2018}]{2018MNRAS.474.2871R}
{Rorai} A.,  {Carswell} R.~F.,  {Haehnelt} M.~G.,  {Becker} G.~D.,  {Bolton}
  J.~S.,   {Murphy} M.~T.,  2018, \mn@doi [\mnras] {10.1093/mnras/stx2862},
  \href {https://ui.adsabs.harvard.edu/abs/2018MNRAS.474.2871R} {474, 2871}

\bibitem[\protect\citeauthoryear{{Rudie} et~al.,}{{Rudie}
  et~al.}{2012a}]{Rudie_12_CGM}
{Rudie} G.~C.,  et~al., 2012a, \mn@doi [\apj] {10.1088/0004-637X/750/1/67},
  \href {https://ui.adsabs.harvard.edu/abs/2012ApJ...750...67R} {750, 67}

\bibitem[\protect\citeauthoryear{{Rudie}, {Steidel}  \& {Pettini}}{{Rudie}
  et~al.}{2012b}]{Rudie_2012_T_rho}
{Rudie} G.~C.,  {Steidel} C.~C.,   {Pettini} M.,  2012b, \mn@doi [\apjl]
  {10.1088/2041-8205/757/2/L30}, \href
  {https://ui.adsabs.harvard.edu/abs/2012ApJ...757L..30R} {757, L30}

\bibitem[\protect\citeauthoryear{{Rudie}, {Steidel}, {Shapley}  \&
  {Pettini}}{{Rudie} et~al.}{2013}]{Rudie_13_CGM}
{Rudie} G.~C.,  {Steidel} C.~C.,  {Shapley} A.~E.,   {Pettini} M.,  2013,
  \mn@doi [\apj] {10.1088/0004-637X/769/2/146}, \href
  {http://cdsads.u-strasbg.fr/abs/2013ApJ...769..146R} {769, 146}

\bibitem[\protect\citeauthoryear{{Schaye}, {Theuns}, {Leonard}  \&
  {Efstathiou}}{{Schaye} et~al.}{1999}]{Schaye_99_IGM_temp}
{Schaye} J.,  {Theuns} T.,  {Leonard} A.,   {Efstathiou} G.,  1999, \mn@doi
  [\mnras] {10.1046/j.1365-8711.1999.02956.x}, \href
  {http://cdsads.u-strasbg.fr/abs/1999MNRAS.310...57S} {310, 57}

\bibitem[\protect\citeauthoryear{{Schaye}, {Theuns}, {Rauch}, {Efstathiou}  \&
  {Sargent}}{{Schaye} et~al.}{2000}]{2000MNRAS.318..817S}
{Schaye} J.,  {Theuns} T.,  {Rauch} M.,  {Efstathiou} G.,   {Sargent} W.~L.~W.,
   2000, \mn@doi [\mnras] {10.1046/j.1365-8711.2000.03815.x}, \href
  {http://adsabs.harvard.edu/abs/2000MNRAS.318..817S} {318, 817}

\bibitem[\protect\citeauthoryear{{Schlickeiser}, {Ibscher}  \&
  {Supsar}}{{Schlickeiser} et~al.}{2012}]{2012ApJ...758..102S}
{Schlickeiser} R.,  {Ibscher} D.,   {Supsar} M.,  2012, \mn@doi [\apj]
  {10.1088/0004-637X/758/2/102}, \href
  {http://adsabs.harvard.edu/abs/2012ApJ...758..102S} {758, 102}

\bibitem[\protect\citeauthoryear{{Schlickeiser}, {Krakau}  \&
  {Supsar}}{{Schlickeiser} et~al.}{2013}]{2013ApJ...777...49S}
{Schlickeiser} R.,  {Krakau} S.,   {Supsar} M.,  2013, \mn@doi [\apj]
  {10.1088/0004-637X/777/1/49}, \href
  {http://adsabs.harvard.edu/abs/2013ApJ...777...49S} {777, 49}

\bibitem[\protect\citeauthoryear{{Shalaby}, {Broderick}, {Chang}, {Pfrommer},
  {Lamberts}  \& {Puchwein}}{{Shalaby} et~al.}{2017a}]{2017ApJ...841...52S}
{Shalaby} M.,  {Broderick} A.~E.,  {Chang} P.,  {Pfrommer} C.,  {Lamberts} A.,
   {Puchwein} E.,  2017a, \mn@doi [\apj] {10.3847/1538-4357/aa6d13}, \href
  {https://ui.adsabs.harvard.edu/abs/2017ApJ...841...52S} {841, 52}

\bibitem[\protect\citeauthoryear{{Shalaby}, {Broderick}, {Chang}, {Pfrommer},
  {Lamberts}  \& {Puchwein}}{{Shalaby} et~al.}{2017b}]{2017ApJ...848...81S}
{Shalaby} M.,  {Broderick} A.~E.,  {Chang} P.,  {Pfrommer} C.,  {Lamberts} A.,
   {Puchwein} E.,  2017b, \mn@doi [\apj] {10.3847/1538-4357/aa8b17}, \href
  {https://ui.adsabs.harvard.edu/abs/2017ApJ...848...81S} {848, 81}

\bibitem[\protect\citeauthoryear{{Shalaby}, {Broderick}, {Chang}, {Pfrommer},
  {Lamberts}  \& {Puchwein}}{{Shalaby} et~al.}{2018}]{2018ApJ...859...45S}
{Shalaby} M.,  {Broderick} A.~E.,  {Chang} P.,  {Pfrommer} C.,  {Lamberts} A.,
   {Puchwein} E.,  2018, \mn@doi [\apj] {10.3847/1538-4357/aabe92}, \href
  {https://ui.adsabs.harvard.edu/abs/2018ApJ...859...45S} {859, 45}

\bibitem[\protect\citeauthoryear{{Shalaby}, {Broderick}, {Chang}, {Pfrommer},
  {Puchwein}  \& {Lamberts}}{{Shalaby} et~al.}{2020}]{2020JPlPh..86b5301S}
{Shalaby} M.,  {Broderick} A.~E.,  {Chang} P.,  {Pfrommer} C.,  {Puchwein} E.,
   {Lamberts} A.,  2020, \mn@doi [Journal of Plasma Physics]
  {10.1017/S0022377820000215}, \href
  {https://ui.adsabs.harvard.edu/abs/2020JPlPh..86b5301S} {86, 535860201}

\bibitem[\protect\citeauthoryear{{Shen} et~al.,}{{Shen}
  et~al.}{2009}]{2009ApJ...697.1656S}
{Shen} Y.,  et~al., 2009, \mn@doi [\apj] {10.1088/0004-637X/697/2/1656}, \href
  {https://ui.adsabs.harvard.edu/abs/2009ApJ...697.1656S} {697, 1656}

\bibitem[\protect\citeauthoryear{{Shull}, {Smith}  \& {Danforth}}{{Shull}
  et~al.}{2012}]{2012ApJ...759...23S}
{Shull} J.~M.,  {Smith} B.~D.,   {Danforth} C.~W.,  2012, \mn@doi [\apj]
  {10.1088/0004-637X/759/1/23}, \href
  {http://adsabs.harvard.edu/abs/2012ApJ...759...23S} {759, 23}

\bibitem[\protect\citeauthoryear{{Sironi} \& {Giannios}}{{Sironi} \&
  {Giannios}}{2014}]{2014ApJ...787...49S}
{Sironi} L.,  {Giannios} D.,  2014, \mn@doi [\apj]
  {10.1088/0004-637X/787/1/49}, \href
  {http://adsabs.harvard.edu/abs/2014ApJ...787...49S} {787, 49}

\bibitem[\protect\citeauthoryear{{Slosar} et~al.,}{{Slosar}
  et~al.}{2013}]{2013JCAP...04..026S}
{Slosar} A.,  et~al., 2013, \mn@doi [JCAP] {10.1088/1475-7516/2013/04/026},
  \href {http://adsabs.harvard.edu/abs/2013JCAP...04..026S} {4, 026}

\bibitem[\protect\citeauthoryear{{Springel}}{{Springel}}{2005}]{2005MNRAS.364.1105S}
{Springel} V.,  2005, \mn@doi [\mnras] {10.1111/j.1365-2966.2005.09655.x},
  \href {http://adsabs.harvard.edu/abs/2005MNRAS.364.1105S} {364, 1105}

\bibitem[\protect\citeauthoryear{{Springel} \& {Hernquist}}{{Springel} \&
  {Hernquist}}{2002}]{2002MNRAS.333..649S}
{Springel} V.,  {Hernquist} L.,  2002, \mn@doi [\mnras]
  {10.1046/j.1365-8711.2002.05445.x}, \href
  {http://adsabs.harvard.edu/abs/2002MNRAS.333..649S} {333, 649}

\bibitem[\protect\citeauthoryear{{Steidel}, {Erb}, {Shapley}, {Pettini},
  {Reddy}, {Bogosavljevi{\'c}}, {Rudie}  \& {Rakic}}{{Steidel}
  et~al.}{2010}]{Steidel_10_KBSS}
{Steidel} C.~C.,  {Erb} D.~K.,  {Shapley} A.~E.,  {Pettini} M.,  {Reddy} N.,
  {Bogosavljevi{\'c}} M.,  {Rudie} G.~C.,   {Rakic} O.,  2010, \mn@doi [\apj]
  {10.1088/0004-637X/717/1/289}, \href
  {https://ui.adsabs.harvard.edu/abs/2010ApJ...717..289S} {717, 289}

\bibitem[\protect\citeauthoryear{{Tiede}, {Broderick}, {Shalaby}, {Pfrommer},
  {Puchwein}, {Chang}  \& {Lamberts}}{{Tiede}
  et~al.}{2017}]{2017ApJ...850..157T}
{Tiede} P.,  {Broderick} A.~E.,  {Shalaby} M.,  {Pfrommer} C.,  {Puchwein} E.,
  {Chang} P.,   {Lamberts} A.,  2017, \mn@doi [\apj]
  {10.3847/1538-4357/aa9375}, \href
  {https://ui.adsabs.harvard.edu/abs/2017ApJ...850..157T} {850, 157}

\bibitem[\protect\citeauthoryear{{Tiede}, {Broderick}, {Shalaby}, {Pfrommer},
  {Puchwein}, {Chang}  \& {Lamberts}}{{Tiede}
  et~al.}{2020}]{2020ApJ...892..123T}
{Tiede} P.,  {Broderick} A.~E.,  {Shalaby} M.,  {Pfrommer} C.,  {Puchwein} E.,
  {Chang} P.,   {Lamberts} A.,  2020, \mn@doi [\apj]
  {10.3847/1538-4357/ab737e}, \href
  {https://ui.adsabs.harvard.edu/abs/2020ApJ...892..123T} {892, 123}

\bibitem[\protect\citeauthoryear{{Upton Sanderbeck} \& {Bird}}{{Upton
  Sanderbeck} \& {Bird}}{2020}]{2020arXiv200205733U}
{Upton Sanderbeck} P.,  {Bird} S.,  2020, arXiv e-prints, \href
  {https://ui.adsabs.harvard.edu/abs/2020arXiv200205733U} {p. arXiv:2002.05733}

\bibitem[\protect\citeauthoryear{{Vafin}, {Rafighi}, {Pohl}  \&
  {Niemiec}}{{Vafin} et~al.}{2018}]{2018ApJ...857...43V}
{Vafin} S.,  {Rafighi} I.,  {Pohl} M.,   {Niemiec} J.,  2018, \mn@doi [\apj]
  {10.3847/1538-4357/aab552}, \href
  {https://ui.adsabs.harvard.edu/abs/2018ApJ...857...43V} {857, 43}

\bibitem[\protect\citeauthoryear{{Viel}, {Haehnelt}  \& {Springel}}{{Viel}
  et~al.}{2004}]{2004MNRAS.354..684V}
{Viel} M.,  {Haehnelt} M.~G.,   {Springel} V.,  2004, \mn@doi [\mnras]
  {10.1111/j.1365-2966.2004.08224.x}, \href
  {http://adsabs.harvard.edu/abs/2004MNRAS.354..684V} {354, 684}

\bibitem[\protect\citeauthoryear{{Viel}, {Bolton}  \& {Haehnelt}}{{Viel}
  et~al.}{2009}]{2009MNRAS.399L..39V}
{Viel} M.,  {Bolton} J.~S.,   {Haehnelt} M.~G.,  2009, \mn@doi [\mnras]
  {10.1111/j.1745-3933.2009.00720.x}, \href
  {http://adsabs.harvard.edu/abs/2009MNRAS.399L..39V} {399, L39}

\bibitem[\protect\citeauthoryear{{Wechsler}, {Bullock}, {Primack}, {Kravtsov}
  \& {Dekel}}{{Wechsler} et~al.}{2002}]{2002ApJ...568...52W}
{Wechsler} R.~H.,  {Bullock} J.~S.,  {Primack} J.~R.,  {Kravtsov} A.~V.,
  {Dekel} A.,  2002, \mn@doi [\apj] {10.1086/338765}, \href
  {https://ui.adsabs.harvard.edu/abs/2002ApJ...568...52W} {568, 52}

\bibitem[\protect\citeauthoryear{{Worseck} et~al.,}{{Worseck}
  et~al.}{2011}]{2011ApJ...733L..24W}
{Worseck} G.,  et~al., 2011, \mn@doi [\apjl] {10.1088/2041-8205/733/2/L24},
  \href {http://adsabs.harvard.edu/abs/2011ApJ...733L..24W} {733, L24}

\bibitem[\protect\citeauthoryear{{Worseck}, {Prochaska}, {Hennawi}  \&
  {McQuinn}}{{Worseck} et~al.}{2016}]{2016ApJ...825..144W}
{Worseck} G.,  {Prochaska} J.~X.,  {Hennawi} J.~F.,   {McQuinn} M.,  2016,
  \mn@doi [\apj] {10.3847/0004-637X/825/2/144}, \href
  {http://adsabs.harvard.edu/abs/2016ApJ...825..144W} {825, 144}

\makeatother
\end{thebibliography}



\appendix

\section{Comparison with Puchwein et al, 2012}
Here we compare the transmitted flux PDF and power spectrum from the simulations presented here with previous work presented in P12. Given the systematic uncertainties which can affect the interpretation of both observational and simulated data (see \S\ref{sec:discussion}) we only compare simulations with each other and do not attempt to compare with observational data here. To allow for an exact comparison, we use outputs at the exact same redshift from both simulation sets and only present the redshift range relevant to the work presented here.

\begin{figure}
\centering
\includegraphics[width = .45\textwidth ]{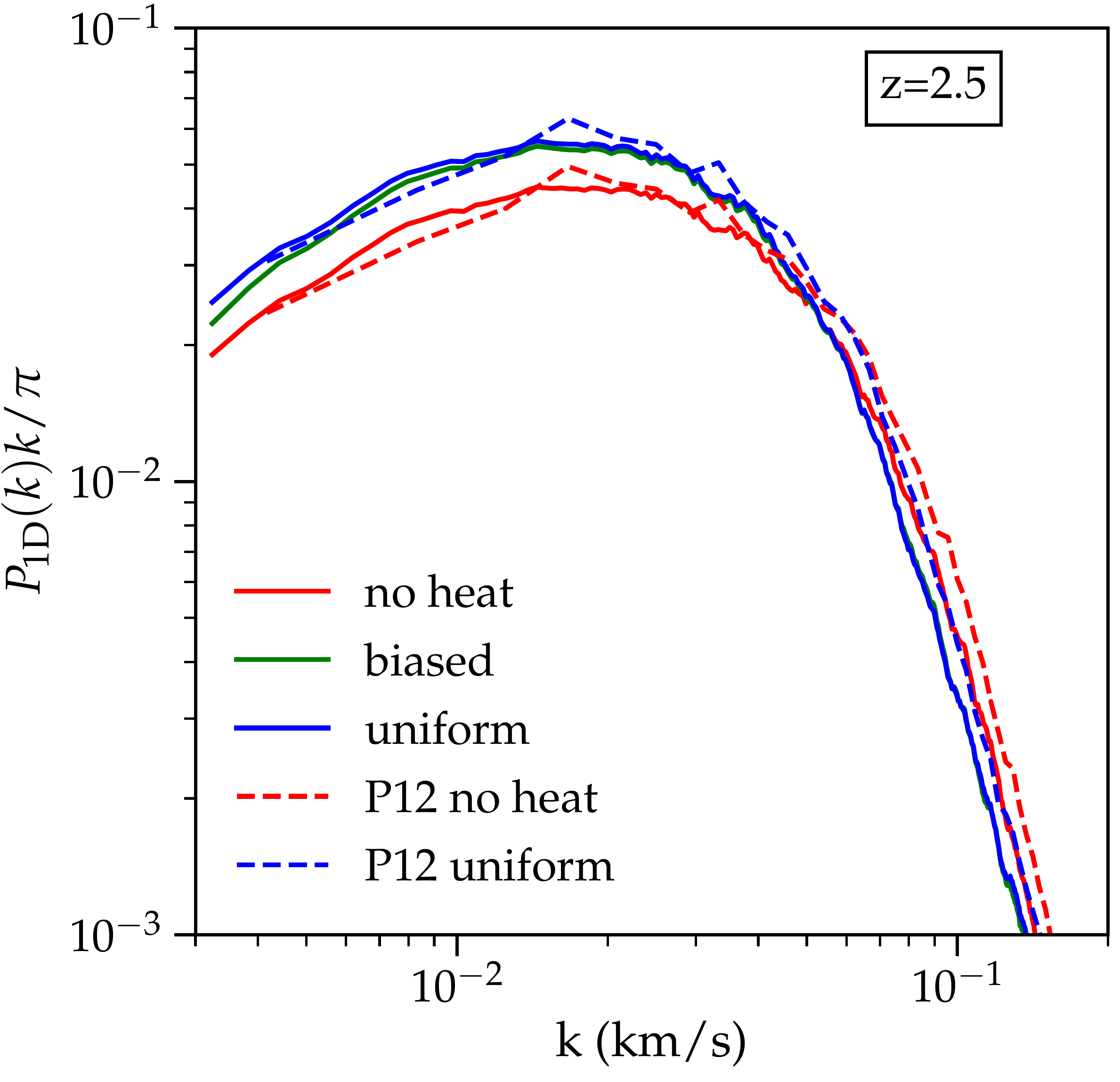}
\caption{Power spectra $P_\mathrm{1D}(k)k/\pi$ of transmitted flux contrast $\exp(-\tau)/\langle\exp(-\tau)\rangle - 1$ in the three heating models presented here (solid lines) and the uniform and unheated models presented in P12 at redshift $z=2.5$(dashed lines). Aside from rescaling to the mean transmitted flux \citep{Becker_13_tau}, no treatment has been applied to the simulations and a comparison with observations is not straightforward.}
\label{fig:powespec}
\end{figure}

\begin{figure*}
\centering
\includegraphics[width = .7\textwidth ]{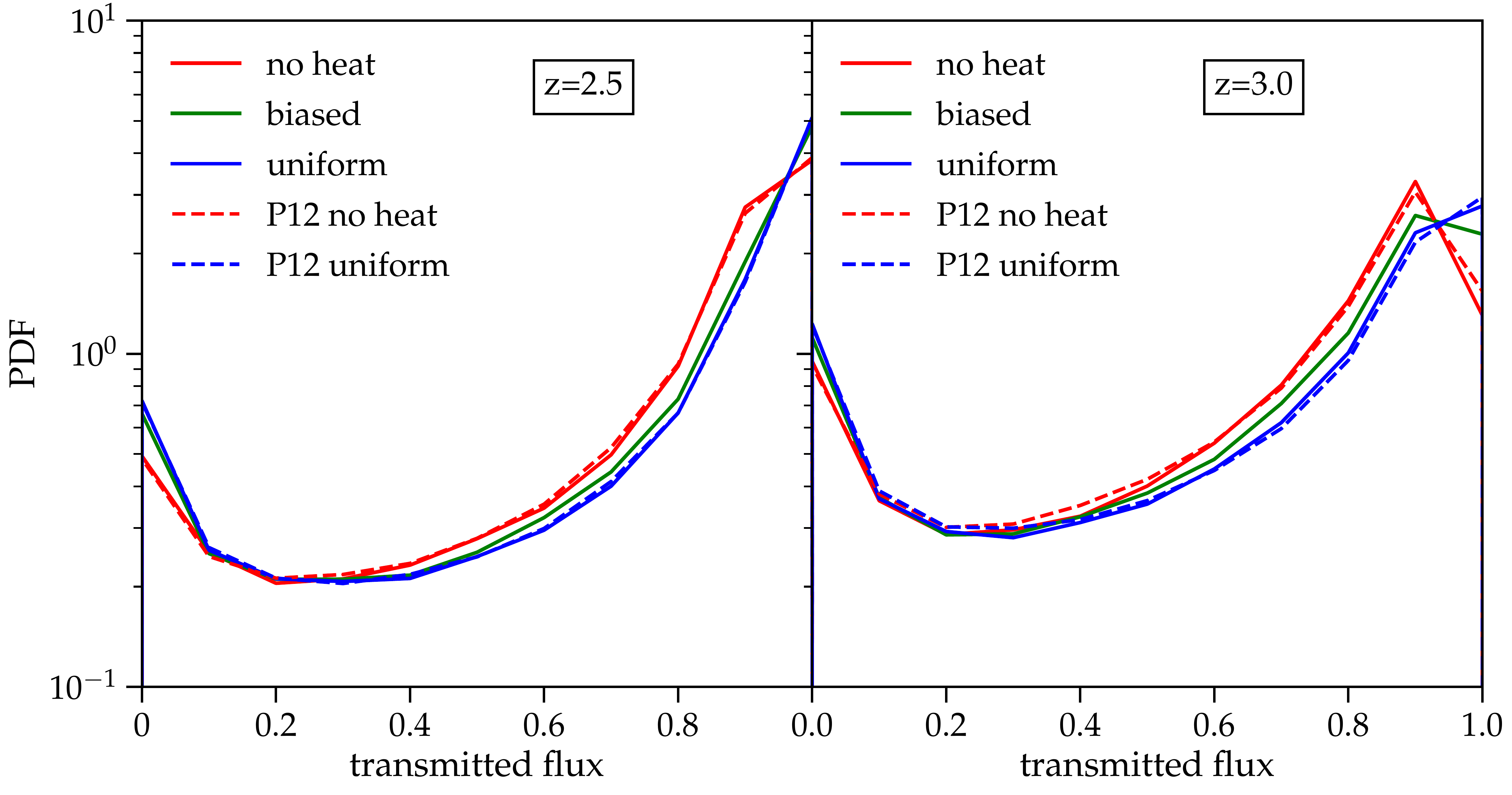}
\caption{Regular ($A=1$) flux PDF in the three heating models presented here (solid lines) and the uniform and unheated model presented in P12 at redshifts $z=2.5$ and 3. Aside from rescaling to the mean transmitted flux \citep{Becker_13_tau}, no treatment has been applied to the simulations and a comparison with observations is not straightforward.}
\label{fig:fluxPDF}
\end{figure*}

Figure~\ref{fig:powespec} shows the one-dimensional power spectrum of the Ly$\alpha$ forest in the simulations, similarly to the right panel in Fig.~8 of P12. Both sets of simulations show good agreement for the different models although slightly more power is present at small scales in P12. This is likely due to the higher resolution of the P12 simulations. Figure~\ref{fig:fluxPDF} shows the transmitted flux PDF in both sets of simulations, similarly to the right panel in Fig.~7 of P12. Both sets show a very good agreement.

These comparisons show that the simulations presented here are very consistent with those presented in P12. However, this work and P12 lead to different conclusions regarding the blazar heating of the low density IGM. This emphasizes the need to use more sensitive observational constraints (such as the linewidth and column density distributions or rescaled flux PDF) and to consider a wide redshift range to understand the thermal evolution of the IGM.

\section{Predicted rescaled flux PDF}

Table~\ref{tab:predict_PDF} shows the rescaled (A=10) flux pdf between $z=2-2/5$ predicted by various heating models.
\begin{table*}
	\centering
	\begin{tabular}{c|cc|cc|cc}

   bin center & PDF$_{\mathrm{noheat}}$ & $\sigma_{\mathrm{noheat}}$ & PDF$_{\mathrm{biased}}$& $\sigma_{\mathrm{biased}}$& PDF$_{\mathrm{uniform}}$ & $\sigma_{\mathrm{uniform}}$ \\ 
		\hline

 -9$\times$10$^{-3}$ &4.47&  0.31    & 4.77 & 0.326 &  4.75 &  0.346 \\
0.05 & 1.11&  0.09 & 0.92 & 0.078 & 0.85 & 0.084 \\
0.11 & 0.73 &  0.055 & 0.57 & 0.051 & 0.54 & 0.052 \\
0.17 & 0.64 &  0.059 & 0.48 & 0.049 & 0.45 & 0.044 \\
0.23 & 0.6&  0.051 & 0.45 & 0.041 & 0.42 &  0.041 \\
0.29 & 0.59&  0.053 & 0.45 &  0.045 & 0.41 &  0.036 \\
0.35 & 0.59&  0.05 & 0.45 &  0.046 & 0.42 &  0.04  \\
0.41 &  0.61 &  0.051& 0.47 & 0.046 & 0.43 & 0.042\\
0.47 & 0.64 &  0.051& 0.49 &  0.044& 0.46 & 0.037  \\
0.52 &  0.66 &  0.051 & 0.53 &  0.045 & 0.52 & 0.044 \\
0.60 & 0.69&  0.048& 0.6 & 0.047 &  0.58 & 0.047  \\
0.66 & 0.74 & 0.057 &  0.69 & 0.053 & 0.68 & 0.053  \\
0.72 & 0.79 & 0.061 & 0.81& 0.056 & 0.82 &  0.057\\
0.77 &  0.82& 0.065 & 0.97& 0.067 & 1.02 & 0.075\\  
0.84 &  0.8 & 0.071 & 1.1 & 0.085 &  1.20 &  0.110\\
0.90 &  0.74 &  0.063 & 1.09 & 0.083 & 1.19 & 0.096\\
0.96 &  0.6 & 0.054 & 0.88 &  0.055&  0.94 &  0.06\\
1.02 & 0.43 & 0.032 & 0.55 & 0.025 &  0.57 & 0.027 \\
	\hline
	\end{tabular}
	
    \caption{Numerical values for Fig.~\ref{fig:PDF_predict} presenting the rescaled flux PDF (A=10) for $2 <z\leq 2.5$ and its standard deviation in the three heating models.}
	\label{tab:predict_PDF}
\end{table*}

\bsp	
\label{lastpage}
\end{document}